\documentclass[twocolumn]{aastex631}

\newcommand{\nangle}[1]{$\left\langle #1 \right\rangle$}

\usepackage{multirow}
\usepackage{pbox}
\shorttitle{CCN}
\shortauthors{Jing et al.}
\graphicspath{{./}{figures/}}
\begin{document}

\title{Criss Cross Nebula: Case study of shock regions with resolved microstructures at scales of $\sim$1000 AU}

\correspondingauthor{Tao Jing}
\email{jingt20@mails.tsinghua.edu.cn}
\correspondingauthor{Cheng Li}
\email{cli2015@tsinghua.edu.cn}

\author[0009-0004-6271-4321]{Tao Jing}
\affiliation{Department of Astronomy, Tsinghua University, Beijing 100084, China}

\author[0000-0002-8711-8970]{Cheng Li}
\affiliation{Department of Astronomy, Tsinghua University, Beijing 100084, China}

\author{Renbin Yan}
\affiliation{Department of Physics, The Chinese University of Hong Kong, Sha Tin, NT, Hong Kong, China}

\author{Cheng Cheng}
\affiliation{National Astronomical Observatories, 20A Datun Road, Beijing 100101, China}

\author{Wei Zhang}
\affiliation{National Astronomical Observatories, 20A Datun Road, Beijing 100101, China}

\author{Xihan Ji}
\affiliation{Department of Physics and Astronomy, University of Kentucky, 505 Rose St., Lexington, KY 40506-0057, USA}

\author{Niu Li}
\affiliation{Department of Astronomy, Tsinghua University, Beijing 100084, China}

\author{Jing Wang}
\affiliation{Kavli Institute for Astronomy and Astrophysics, Peking University, Beijing 100871, China}

\author{Chaojian Wu}
\affiliation{National Astronomical Observatories, 20A Datun Road, Beijing 100101, China}

\author{Haibo Yuan}
\affiliation{Department of Astronomy, Beijing Normal University, 19 Xinjiekouwai St., Beijing 100875, China}



\begin{abstract}
Using integral field spectroscopy from MaNGA, we study the resolved
microstructures in a shocked region in Criss Cross Nebula (CCN), with
an unprecedentedly high resolution of $\lesssim$1000 AU.  We measure
surface brightness maps for 34 emission lines, which can be broadly
divided into three categories: (1) the [O{\sc iii}] $\lambda$5007-like
group including seven high-ionization
lines and two [O{\sc ii}] auroral lines which uniformly present a
remarkable lane structure, (2) the H$\alpha$ $\lambda$6563-like group including 23
low-ionization or recombination lines which present a clump-like
structure, and (3) [O{\sc ii}] $\lambda$3726 and [O{\sc ii}]
$\lambda$3729 showing high densities at both the [O{\sc iii}]
$\lambda$5007 lane and the H$\alpha$ clump.  We use these measurements
to constrain resolved shock models implemented in {\tt MAPPINGS V}.  We find
our data can be reasonably well-fitted by a model which includes a
plane-parallel shock with a velocity of $133\pm5$ km/s, 
plus an isotropic two-dimensional Gaussian component which 
is likely another clump of gas ionized by photons from the shocked region,
and a constant background. We compare the
electron density and temperature profiles as predicted by our model
with those calculated using observed emission line ratios. We find
different line ratios to provide inconsistent
temperature maps, and the discrepancies can be attributed to
observational effects caused by limited spatial resolution and
projection of the shock geometry, as well as contamination of the
additional Gaussian component. Implications on shock properties and
perspectives on future IFS-based studies of CCN are discussed.
\end{abstract}

%a pre-shock density of $n_{\mathrm{pre}}=1.4\pm0.2$ cm$^{-3}$, a transverse magnetic field of $B=1.6\pm0.3$ $\mu$G and an inclination angle of $19\pm14$ deg, 

%  



\section{Introduction} \label{sec:intro}

Shocks may be produced by multiple processes that are key in galaxy 
formation and evolution, such as stellar winds, supernova feedback, 
AGN feedback, and gas accretion/outflow at different scales. There have 
been many studies of shocks and shock-induced structures
in both the Milky Way and external galaxies, e.g.  
shock-induced filaments in the Orion-Eridanus superbubble \citep{2014MNRAS.441.1095P}, the shock-excited Herbig-Haro objects \citep{Dopita2017_HHObject}, and shock ionization in passive galaxies \citep{2018MNRAS.481..476Y}, star forming galaxies \citep{Shopbell1998, Veilleux2005, Sharp2010, Westmoquette2011, Ho2014}, galaxy mergers \citep{Farage2010, Soto2012A, 2012SotoB, Medling2021}, AGN host galaxies \citep{Dopita1997, Dopita2015, Terao2016, Molina2018, Molina2021, Holden2023} and post starburst galaxies \citep{Alatalo2016}. These studies 
have deepened our understanding of shocks in different aspects. 
However, previous observations of shocked regions are mostly limited 
to either narrow-band imaging of emission lines with very low spectral 
resolution and no kinematics information, or spectroscopy with limited 
spatial coverage and resolution. It is crucial to have resolved 
spectroscopy with high spatial and spectral resolutions in order to fully
understand not only the shocks themselves but also their roles in the
dynamic heating of interstellar medium (ISM) and destruction of interstellar 
clouds.

There have also been many studies of nearby shocked regions, 
such as Crab Nebula at $\sim 2\ \mathrm{kpc}$ 
\citep{Crab_outer, Crab_SNL, Crab1, Crab_progenitor, 
	Crab_ejecta, Crab2, Crab_CFHT}  and IC443 at $\sim 2\ \mathrm{kpc}$
\citep{IC443_1, IC443_2, IC443_3, IC443_distance, IC443_4, IC443_5, 
	IC443_SITELLE, IC443_DYW}, which usually present regular 
	shell-like structures and have high shock velocities and electron densities.
In this work we study the Criss Cross Nebula (CCN), which has 
different structures and physical conditions, compared with 
other well studied nearby shock regions.
Located at ($\alpha = 62.55$ deg, $\delta$ = -4.98 deg) or ($l$ = 196.97 deg, 
$b$ = -37.83 deg) and with an angular size of $6^\prime\times3^\prime$, 
CCN is dubbed by its unique morphology and diverse
microstructures and is believed to be produced by one or multiple shocks
(see \autoref{fig:large_scale_image}). This nebula was firstly studied by 
\citet[][hereafter Z97]{Z97} and \citet[][hereafter Z99]{Z99}, who suggested that the CCN is ionized by a 
shock with a velocity of about 40 km/s and electron density in the range 
$1 - 100 \ \mathrm{cm}^{-3}$ based on narrow-band images and long-slit 
spectra. A following study by \citet[][hereafter T07]{T07} found a photoionized
arc-like structure of H$\alpha$ emission near the CCN, which was called 
``canopy'' by the authors considering its shape and the relative position to
the CCN. Though still uncertain, the distance to CCN should be within 
$\sim400$ pc and could be even more nearby, at $\sim150$ pc as 
argued in T07. CCN is projected at the same area of the Orion-Eridanus superbubble and is very close to Arc A, a major structure within the bubble. 
The small distance, cross-like morphology, low shock velocity, and
complex large-scale environment make the CCN a unique object for 
studies of shocks.

In this work we analyze the integral field spectroscopy (IFS) data for a small region
of the CCN ($\sim20^{\prime\prime}\times20^{\prime\prime}$), which was 
observed as a filler target by the Mapping Nearby Galaxies 
at Apache Point Observatory \citep[MaNGA;][]{MaNGA_Survey} survey. 
At a distance of 150-400 pc, the spatial resolution $\sim2.5^{\prime\prime}$ 
of MaNGA corresponds to a physical scale of $\sim375\ \mathrm{AU}$ 
to $1000\ \mathrm{AU}$. Although covering only a small part of the nebula,
the IFS data from MaNGA allows for the first time a case study of resolved
microstructures in shock regions with unprecedentedly-high resolution. 
We will measure all the emission lines in the MaNGA FoV and use the 
measurements to constrain both physical parameters and geometry parameters 
of the underlying shock.

The paper is structured as follows. In \autoref{sec:data} we describe our data
and measure the emission lines. In \autoref{sec:s_model_res} we use the 
emission line measurements to constrain shock models. We discuss in  
\autoref{sec:discussion} and summarize our work in \autoref{sec:summary}.

\section{Data and Measurements} \label{sec:data}

\subsection{MaNGA Data} \label{sec:manga_data}

\begin{figure*}
	\centering
	\includegraphics[width=\textwidth]{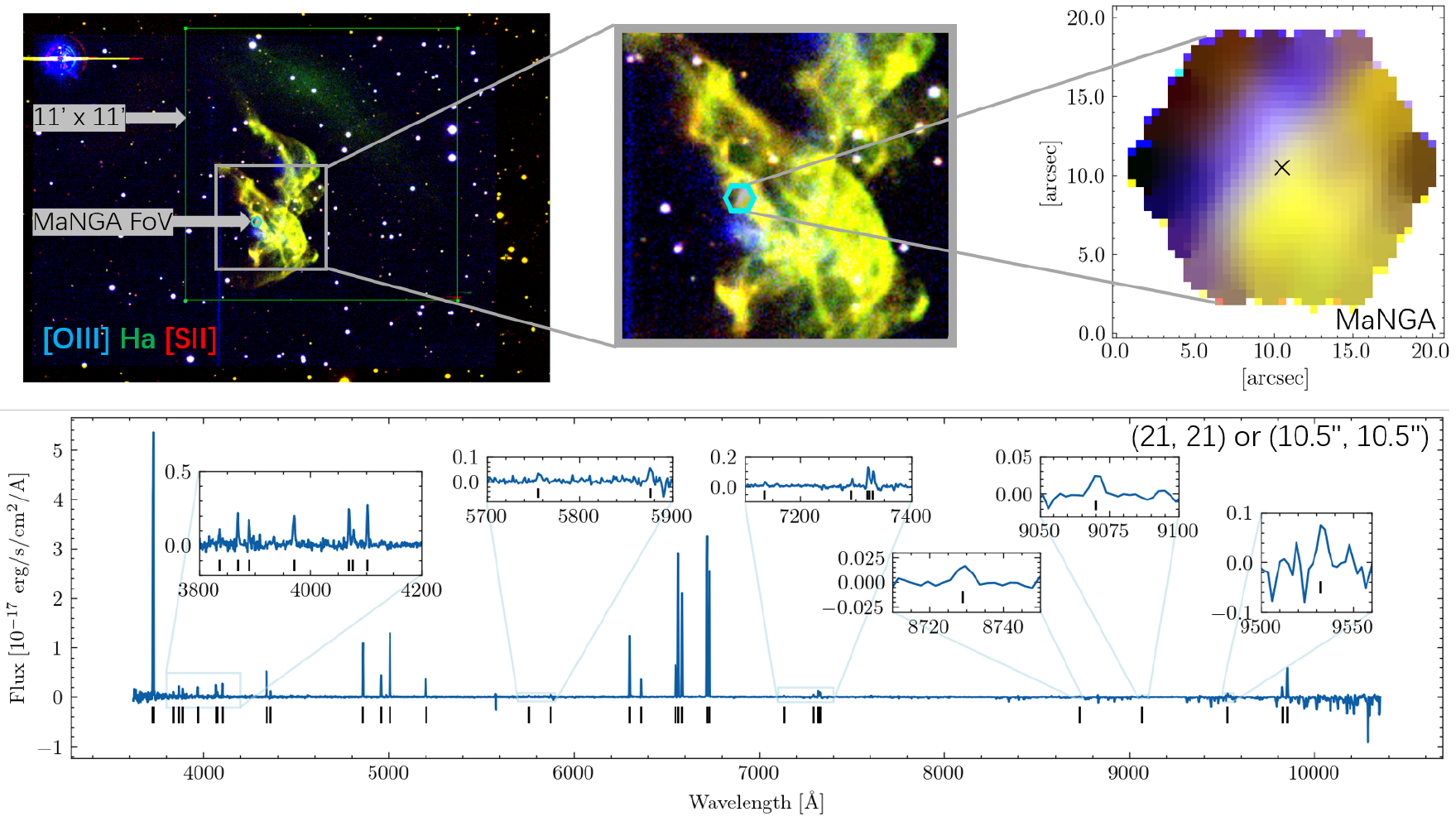}
	\caption{{\it Upper:} Criss-Cross Nebula (CCN) as seen at different scales. The left two images are constructed using narrowband images of [S{\sc ii}] $\lambda\lambda$6717, 6731 and H$\alpha$ $\lambda$6563 from \cite{T07} and [O{\sc iii}] $\lambda \lambda$4959, 5007 obtained in this work. The rightmost image is based on MaNGA data of the same lines. {\it Lower:} the MaNGA spectrum of the central spaxel in the MaNGA field of view (marked as a black cross in the upper-right image). The short black lines below the spectrum indicate the position of 34 emission lines that are measured and studied in this work. Some of the lines are more closely shown in the insets.}
	\label{fig:large_scale_image}
\end{figure*}

A small part of the CCN has been observed as a filler target by the MaNGA survey, 
which is one of the three major experiments of the fourth generation Sloan Digital Sky Survey 
\citep[SDSS-IV;][]{SDSSIV}. During its six-year operation from July 2014 through August 2020, 
MaNGA obtained integral-field spectroscopy (IFS) for $\ga 10^4$ nearby galaxies selected from 
the NASA Sloan Atlas \citep[NSA;][]{NSA2011}. The IFS data were obtained using 17 science 
integral-field units (IFUs) with a field of view (FoV) ranging from $12''$ to $32''$
\citep{MaNGA_Fiber}. The MaNGA spectra
cover a wavelength coverage from 3600$\text{\AA}$ to 10300$\text{\AA}$ with a spectral resolution 
of $R \sim 2000$, taken with the two dual-channel BOSS spectrographs \citep{2013AJ....146...32S} 
on the Sloan 2.5-meter telescope at the Apache Point Observatory \citep{2006AJ....131.2332G}. 
Details of MaNGA sample selection, flux calibration,  survey execution and observing strategy 
can be found in \cite{2017AJ....154...86W}, \cite{law2015}, and \cite{2016AJ....152..197Y, 2016AJ....151....8Y}. 
MaNGA raw data were reduced using the MaNGA data reduction pipeline \citep[DRP;][]{MaNGA_DRP}
which produces a 3D datacube for each target with an effective spatial resolution that can be 
described by a Gaussian with a full width at half maximum (FWHM) of $\sim2.5^{\prime\prime}$.
All the MaNGA data, including raw data and DRP products, are released as part of the 
final data release of SDSS-IV
\citep[DR17\footnote{\url{https://www.sdss.org/dr17/manga/}};][]{SDSS_DR17}.

We use the datacube produced by the MaNGA DRP for this study. The upper-right panel in 
\autoref{fig:large_scale_image} shows the MaNGA FoV, a hexagon covering approximately 
$20'' \times 20''$, thus only a small part of the CCN as indicated by the cyan hexagon in the 
upper-left and upper-middle panels. As an example, the lower panel of the same figure shows the 
spectrum of the central pixel (indicated by the black cross in the upper-right panel), with 
the short black lines marking some of the emission lines.

The MaNGA DRP creates a sky model for each plate based on data from the sky fibers, 
in a way that is designed and optimized for nearby galaxies. For the CCN, however, 
the ``background'' measured by the sky fibers is contaminated by the emission from 
diffuse ionized gas surrounding the CCN. We apply our spectral fitting method described in 
Section \ref{sec:emission_line_measurements} to the sky model constructed by DRP, 
finding some emission lines to be over-subtracted. We do not attempt to rebuild the 
sky model. Rather, we measure the flux of these over-subtracted lines in the sky 
model, and we take into account their effect in the following analysis. A total of 4 
emission lines are confirmed in the sky model including [O{\sc ii}] $\lambda\lambda$3726, 3729 and [S{\sc ii}] $\lambda\lambda$6717, 6731, 
whose redshifts are close to what we measured from the DRP datacube. 
We notice that the over-subtracted flux is typically one order of magnitude 
smaller than the flux measured from the DRP spectrum, and so the flux correction
is generally a tiny effect on our results.

\subsection{Ancillary data} \label{sec:aux_data}

We have obtained the [O{\sc iii}] $\lambda \lambda$4959, 5007 narrowband image for 
the whole CCN,
using the China Near Earth Object Survey Telescope at Xuyi Observational Station
operated by Purple Mountain Observatory (PMO). The observation was taken on December 30, 
2021, with a total exposure time of 1.5 hours. In addition, kindly provided by Bringfried Stecklum,
the narrowband images of [S{\sc ii}] $\lambda\lambda$6717, 6731 and H$\alpha$ $\lambda$6563 
used in T07 are also available for our study. The upper-left panel 
in \autoref{fig:large_scale_image} shows the RGB image of the whole CCN, constructed 
by combining the narrowband images of [O{\sc iii}], H$\alpha$ and [S{\sc ii}]. 

We observed the $^{12}$CO(1-0) line for the CCN using the PMO 13.7-m millimeter 
telescope \citep{PMO_telescope} located in Delingha, China, over the period from April 26, 2022 through 
May 6, 2022. A total of 136 scans were taken, and each scan was reduced to 
produce a datacube with a pixel size of 30$^{\prime\prime}$ by applying the standard 
reduction pipeline of this telescope. After fitting and subtracting the baseline in the spectrum and masking any heavily affected spectra, all unmasked spectra were stacked to create a final spectrum, which has a rms of $T_\mathrm{CO} = 0.039\ \mathrm{K}$ in $0.15\ \mathrm{km/s}$. This spectrum provides no detection of $^{12}$CO(1-0) with a $3\sigma$ upper limit of $0.14\ \mathrm{K\ km/s}$ when assuming a line width of $10\ \mathrm{km/s}$. If we adopt a conversion factor of $\alpha_{\mathrm{CO}} = 4.35\ (M_\odot/\mathrm{pc}^2)/(\mathrm{K\ km/s})$, this upper limit corresponds to an H$_2$ surface density lower than $0.6\ M_\odot/\mathrm{pc}^2$.

\subsection{Emission line measurements} \label{sec:emission_line_measurements}

The MaNGA spectra present numerous emission lines but almost no 
continuum over the whole wavelength coverage, as can be seen from 
the example spectrum in \autoref{fig:large_scale_image}. Therefore, 
for each spaxel we do not attempt to obtain an overall continuum 
component, and we directly apply our emission line fitting code to 
measure all the detected lines. These include a total of 34 emission
lines, as listed in \autoref{tab:lines}. Although the continuum is close 
to zero in all the spectra, the baseline beside each emission line 
always deviates to some degree from a zero horizontal line due to 
the residual in sky subtraction and/or flux calibration in the MaNGA 
DRP. We determine a local baseline with a linear fitting of the spectrum 
over both sides of each emission line, excluding obvious emission 
features and outliers by sigma clipping.  

\begin{figure*}
	\centering
	\includegraphics[width=\textwidth]{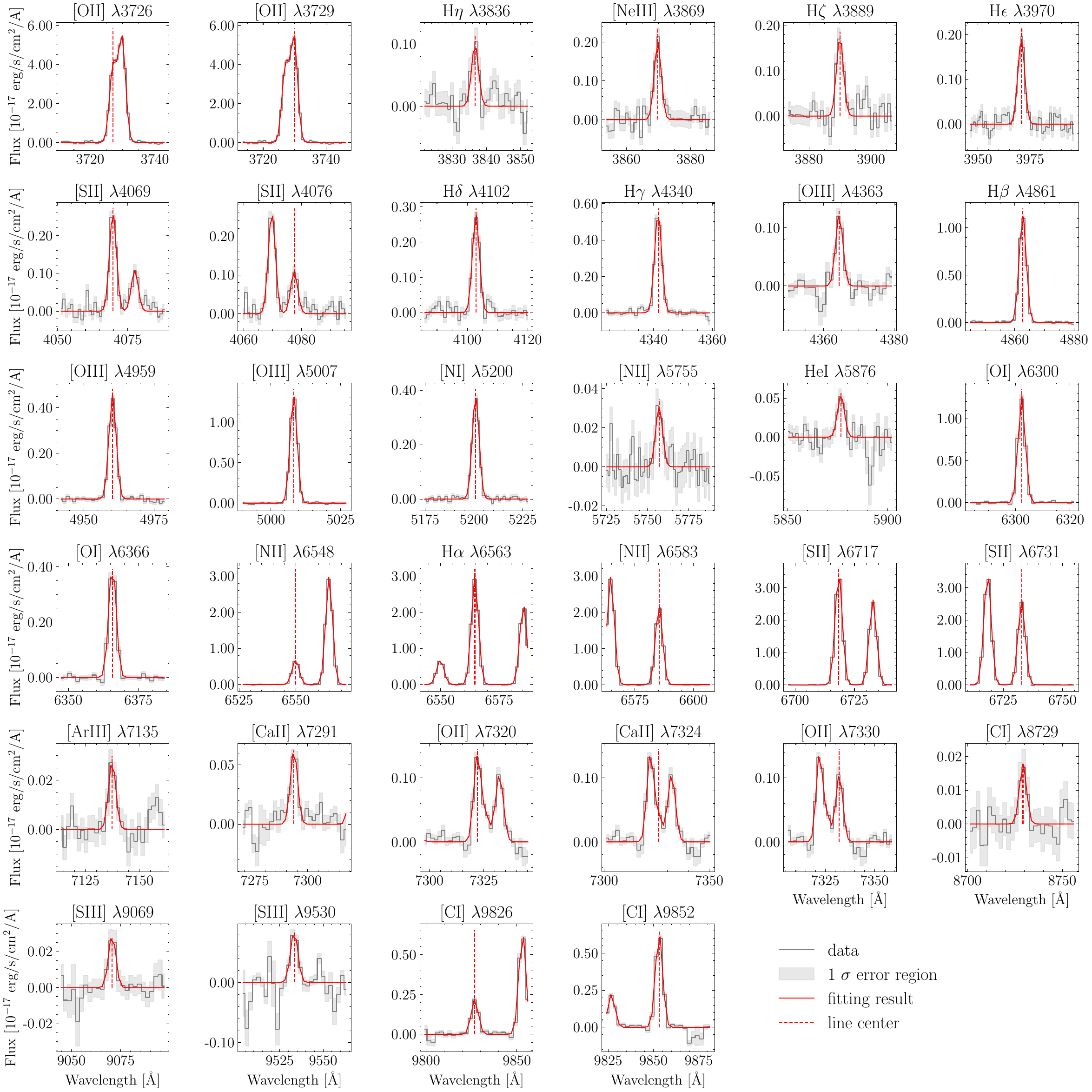}
	\caption{Fitting results for the 34 emission lines of one example spaxel. The grey histogram is the observed spectrum with baseline subtracted,
		and the shaded grey region is the 1 $\sigma$ error.	 The solid red line is the best-fit model, with the line center marked by the dashed vertical line.}
	\label{fig:lines_fitting_21_21}
\end{figure*}

Each emission line is modeled with a single Gaussian. Most of the 
lines are well separated from each other, and each line is fitted 
independently. In some special cases as listed in \autoref{tab:lines}, two or more neighboring lines are fitted simultaneously. In these cases some Gaussian function parameters may be 
tied for reasonable considerations, which are also listed and explained 
in the table. For a given line or a group of lines, the best-fit model 
is determined by minimizing the $\chi^2$, defined as
\begin{equation}
\chi^2 = \sum_i^n \left(\frac{\mathrm{flux}_i - \mathrm{model}_i}{\mathrm{err}_i}\right)^2,
\end{equation} 
where $n$ is the number of wavelength pixels over the emission line 
fitting window centered at the restframe wavelength of the line,
$\mathrm{flux}_i$ and $\mathrm{model}_i$ are the observed and model 
fluxes of the $i$-th pixel,  and $\mathrm{err}_i$ is the error of the 
observed flux of the $i$-th pixel. The width of the fitting window 
ranges from 20 $\text{\AA}$ to 100 $\text{\AA}$, mainly depending on the number 
of lines included in the fitting. As an example, \autoref{fig:lines_fitting_21_21}
displays the result of the fitting for all the 34 lines for the central spaxel 
in the MaNGA datacube. As can be seen, all the lines are well-fitted. In fact, 
the majority of the lines in the whole MaNGA datacube are well-fitted, 
except that for some weak lines such as [Ar{\sc iii}] $\lambda$7135, 
the emission line signal in some spaxels is overwhelmed by noise. 
In such cases, our fitting code gives best-fit models with small fluxes 
and relatively large uncertainties, which can be identified as non-detections
given a signal-to-noise (S/N) threshold. 

\begin{deluxetable*}{lccccc}
	\tablecaption{All the lines fitted in MaNGA data \label{tab:lines}}
	\tablewidth{0pt}
	\tablehead{
		\colhead{Line} & \colhead{\pbox{3cm}{\centering Center wavelength \tablenotemark{a}}} & \colhead{Model \tablenotemark{b}}  & \colhead{Parameter tying} & \colhead{Notes} & \colhead{Morphological group \tablenotemark{c}}\\
		\colhead{} & \colhead{$[\mathrm{\AA}]$} & \colhead{} & \colhead{} & \colhead{} & \colhead{}
	}
	% \decimalcolnumbers
	\startdata
	$\mathrm{[OII]}$ $\lambda$3726 & 3727.1 & \multirow{2}{*}{\pbox{5cm}{\centering 
			$\sum_{i=1}^{i=2}$\cal{G}$_{i}(a_i,\lambda_i,\sigma_i)$}} & \multirow{2}{*}{\pbox{5cm}{\centering $\sigma_1=\sigma_2$}} & \multirow{2}{*}{\tablenotemark{d}} & \nangle{3} \\
	$\mathrm{[OII]}$ $\lambda$3729 & 3729.9 & ~ & & & \nangle{3} \\
	\hline
	H$\eta$ $\lambda$3836 & 3836.5 & \cal{G}$_1(a_1,\lambda_1,\sigma_1)$ & & & \nangle{2} \\
	$\mathrm{[NeIII]}$ $\lambda$3869 & 3869.9 & \cal{G}$_1(a_1,\lambda_1,\sigma_1)$ & & & \nangle{1} \\
	H$\zeta$ $\lambda$3889 & 3890.2 & \cal{G}$_1(a_1,\lambda_1,\sigma_1)$ & & & \nangle{2} \\
	H$\epsilon$ $\lambda$3970 & 3971.2 & \cal{G}$_1(a_1,\lambda_1,\sigma_1)$ & & & \nangle{2} \\
	\hline
	$\mathrm{[SII]}$ $\lambda$4069 & 4069.7 & \multirow{3}{*}{\pbox{5cm}{\centering 
			$\sum_{i=1}^{i=3}$\cal{G}$_{i}(a_i,\lambda_i,\sigma_i)$}} &  \multirow{3}{*}{\pbox{6cm}{\centering $\sigma_1=\sigma_2$}} & \multirow{3}{*}{\tablenotemark{e}} & \nangle{2}\\
	$\mathrm{[SII]}$ $\lambda$4076 & 4077.6 & ~ &  ~ & & \nangle{2}\\
	H$\delta$ $\lambda$4102 & 4102.9 & ~ &  ~ & & \nangle{2}\\
	\hline
	H$\gamma$ $\lambda$4340 & 4341.7 & \cal{G}$_1(a_1,\lambda_1,\sigma_1)$ & & & \nangle{2} \\
	$\mathrm{[OIII]}$ $\lambda$4363 & 4364.4 & \cal{G}$_1(a_1,\lambda_1,\sigma_1)$ & & & \nangle{1} \\
	H$\beta$ $\lambda$4861 & 4862.8 & \cal{G}$_1(a_1,\lambda_1,\sigma_1)$ & & & \nangle{2} \\
	$\mathrm{[OIII]}$ $\lambda$4959 & 4960.3 & \cal{G}$_1(a_1,\lambda_1,\sigma_1)$ & & & \nangle{1} \\
	$\mathrm{[OIII]}$ $\lambda$5007 & 5008.3 & \cal{G}$_1(a_1,\lambda_1,\sigma_1)$ & & & \nangle{1} \\
	$\mathrm{[NI]}$ $\lambda$5200 & 5201.4 & \cal{G}$_1(a_1,\lambda_1,\sigma_1)$ & & & \nangle{2} \\
	$\mathrm{[NII]}$ $\lambda$5755 & 5756.2 & \cal{G}$_1(a_1,\lambda_1,\sigma_1)$ & & & \nangle{2} \\
	HeI $\lambda$5876 & 5877.2 & \cal{G}$_1(a_1,\lambda_1,\sigma_1)$ & & & \nangle{2} \\
	$\mathrm{[OI]}$ $\lambda$6300 & 6302.0 & \cal{G}$_1(a_1,\lambda_1,\sigma_1)$ & & & \nangle{2} \\
	$\mathrm{[OI]}$ $\lambda$6366 & 6368.1 & \cal{G}$_1(a_1,\lambda_1,\sigma_1)$ & & & \nangle{2} \\
	$\mathrm{[NII]}$ $\lambda$6548 & 6549.9 & \cal{G}$_1(a_1,\lambda_1,\sigma_1)$ & & & \nangle{2} \\
	H$\alpha$ $\lambda$6563 & 6564.6 & \cal{G}$_1(a_1,\lambda_1,\sigma_1)$ & & & \nangle{2} \\
	$\mathrm{[NII]}$ $\lambda$6583 & 6585.3 & \cal{G}$_1(a_1,\lambda_1,\sigma_1)$ & & & \nangle{2} \\
	$\mathrm{[SII]}$ $\lambda$6717 & 6718.3 & \cal{G}$_1(a_1,\lambda_1,\sigma_1)$ & & & \nangle{2} \\
	$\mathrm{[SII]}$ $\lambda$6731 & 6732.7 & \cal{G}$_1(a_1,\lambda_1,\sigma_1)$ & & & \nangle{2} \\
	$\mathrm{[ArIII]}$ $\lambda$7135 & 7137.8 & \cal{G}$_1(a_1,\lambda_1,\sigma_1)$ & & & \nangle{1} \\
	\hline
	$\mathrm{[CaII]}$ $\lambda$7291 & 7293.5 & \multirow{4}{*}{\pbox{8cm}{\centering 
			$\sum_{i=1}^{i=4}$\cal{G}$_{i}(a_i,\lambda_i,\sigma_i)$}} & ~ & \multirow{4}{*}{\tablenotemark{f}} & \nangle{2}\\
	$\mathrm{[OII]}$ $\lambda$7320 & 7322.0 & ~ & $\sigma_1=\sigma_3$ & & \nangle{1}\\
	$\mathrm{[CaII]}$ $\lambda$7324 & 7325.9 & ~ & $0.8<\sigma_4/\sigma_2<1.25$ & & \nangle{2}\\
	$\mathrm{[OII]}$ $\lambda$7330 & 7332.0 & ~ & ~ & & \nangle{1}\\
	\hline
	$\mathrm{[CI]}$ $\lambda$8729 & 8729.5 & \cal{G}$_1(a_1,\lambda_1,\sigma_1)$ & & & \nangle{2} \\
	$\mathrm{[SIII]}$ $\lambda$9069 & 9071.4 & \cal{G}$_1(a_1,\lambda_1,\sigma_1)$ & & & \nangle{1} \\
	$\mathrm{[SIII]}$ $\lambda$9530 & 9534.7 & \cal{G}$_1(a_1,\lambda_1,\sigma_1)$ & & & \nangle{1} \\
	\hline
	$\mathrm{[CI]}$ $\lambda$9826 & 9827.0 & \multirow{2}{*}{\pbox{5cm}{\centering
%			\cal{G}$_1(a_1,\lambda_1,\sigma_1)+$\cal{G}$_2(a_2,\lambda_2,\sigma_2)$}} 
   $\sum_{i=1}^{i=2}$\cal{G}$_{i}(a_i,\lambda_i,\sigma_i)$}}
	& \multirow{2}{*}{\pbox{5cm}{\centering $\sigma_1=\sigma_2$}} & \multirow{2}{*}{\tablenotemark{g}} & \nangle{2} \\
	$\mathrm{[CI]}$ $\lambda$9852 & 9853.0 & ~ & ~ & & \nangle{2} \\
	\enddata
	\tablenotetext{a}{Rest-frame vacuum center wavelength for given emission line.}
	\tablenotetext{b}{Generally each emission line is modeled by a single Gaussian, 
		denoted as \cal{G}$_1(a_1,\lambda_1,\sigma_1)$ with three free parameters: 
		amplitude $a_1$, center wavelength $\lambda_1$, and the line width $\sigma_1$
		resulting from kinematical and instrumental broadening. In some cases, 
		neighbouring lines are fitted simultaneously by a linear combination of multiple 
		Gaussians with some parameters tied during fitting, for the reasons provided 
		in column ``Notes''.}
	\tablenotetext{c}{Classification group according to the surface brightness morphology
		of the emission line as described in detail in \autoref{sec:emission_line_measurements}. 
		We use notation \nangle{1} for the [O{\sc iii}] $\lambda$5007 like group, \nangle{2} 
		for the H$\alpha$ $\lambda$6563 like group, and \nangle{3} for the group of 
		[O{\sc ii}] $\lambda$3726 and [O{\sc ii}] $\lambda$3729.}
	\tablenotetext{d}{These two lines are only marginally resolved under MaNGA resolution.}
	\tablenotetext{e}{It is hard to find a clean left and right window for these lines separately, and [S{\sc ii}] $\lambda$4076 is sometimes too weak.}
	\tablenotetext{f}{This configuration aims to alleviate the confusion effect of [Ca{\sc ii}] $\lambda$7324 on [O{\sc ii}] $\lambda\lambda$7320, 7330.}
	\tablenotetext{g}{This configuration aims for more robust results when [C{\sc ii}] $\lambda$9826 is weak.}
\end{deluxetable*}

\begin{figure*}
	\centering
	\includegraphics[width=\textwidth]{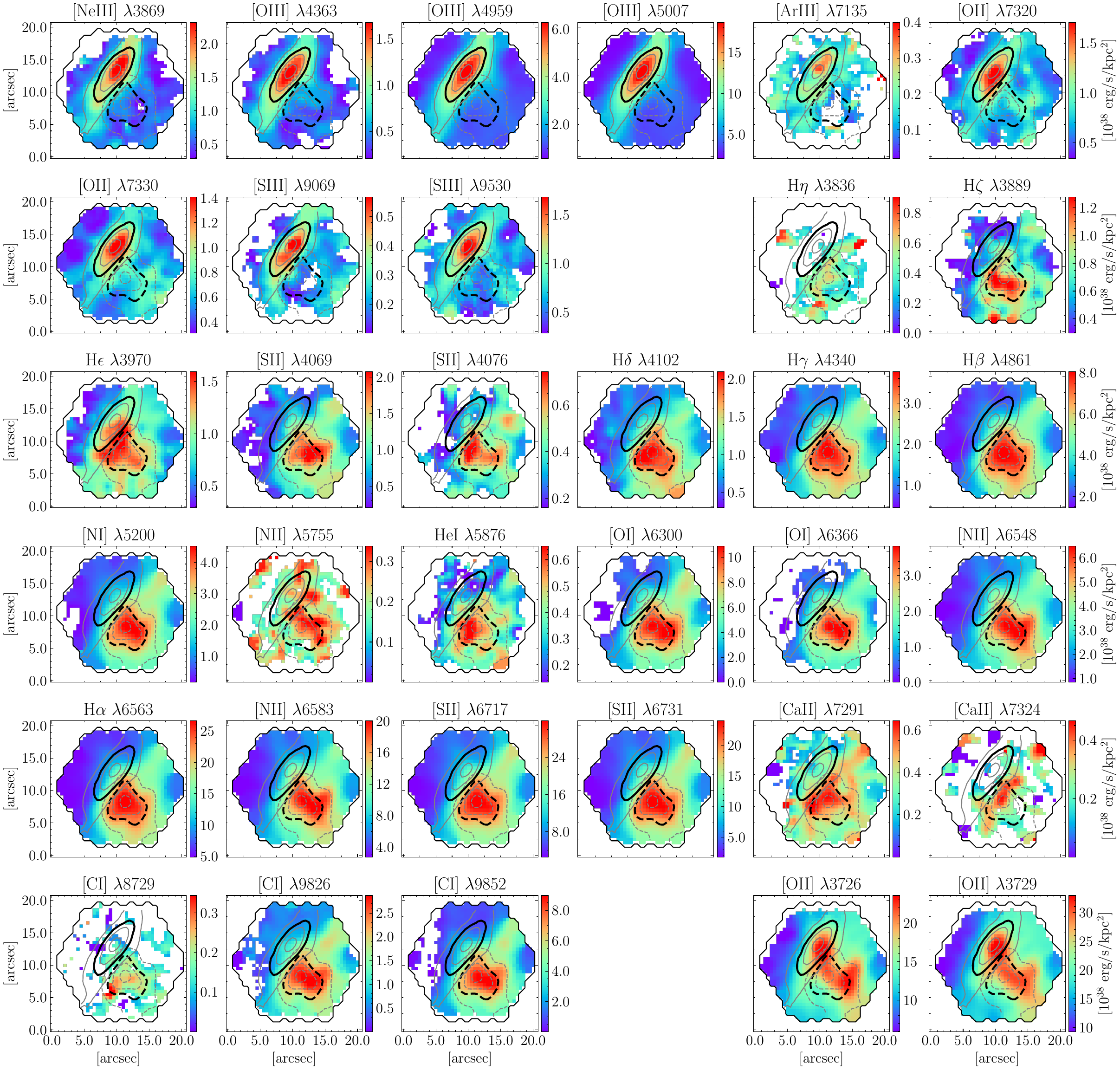}
	\caption{The surface brightness map of all the 34 emission lines measured within the MaNGA FoV in the unit of $10^{38}$ erg/s/kpc$^2$. 
     The solid and dashed gray contours represent the 75\%, 90\% (bolded), 95\%, and 99\% quantiles in the surface brightness 
     map of [O{\sc iii}] $\lambda$5007 and H$\alpha$ $\lambda$6563 lines, respectively, which are repeated in every panel for reference.}
	\label{fig:surface_brightness_all}
\end{figure*}

\begin{figure*}
	\centering
	\includegraphics[width=0.8\textwidth]{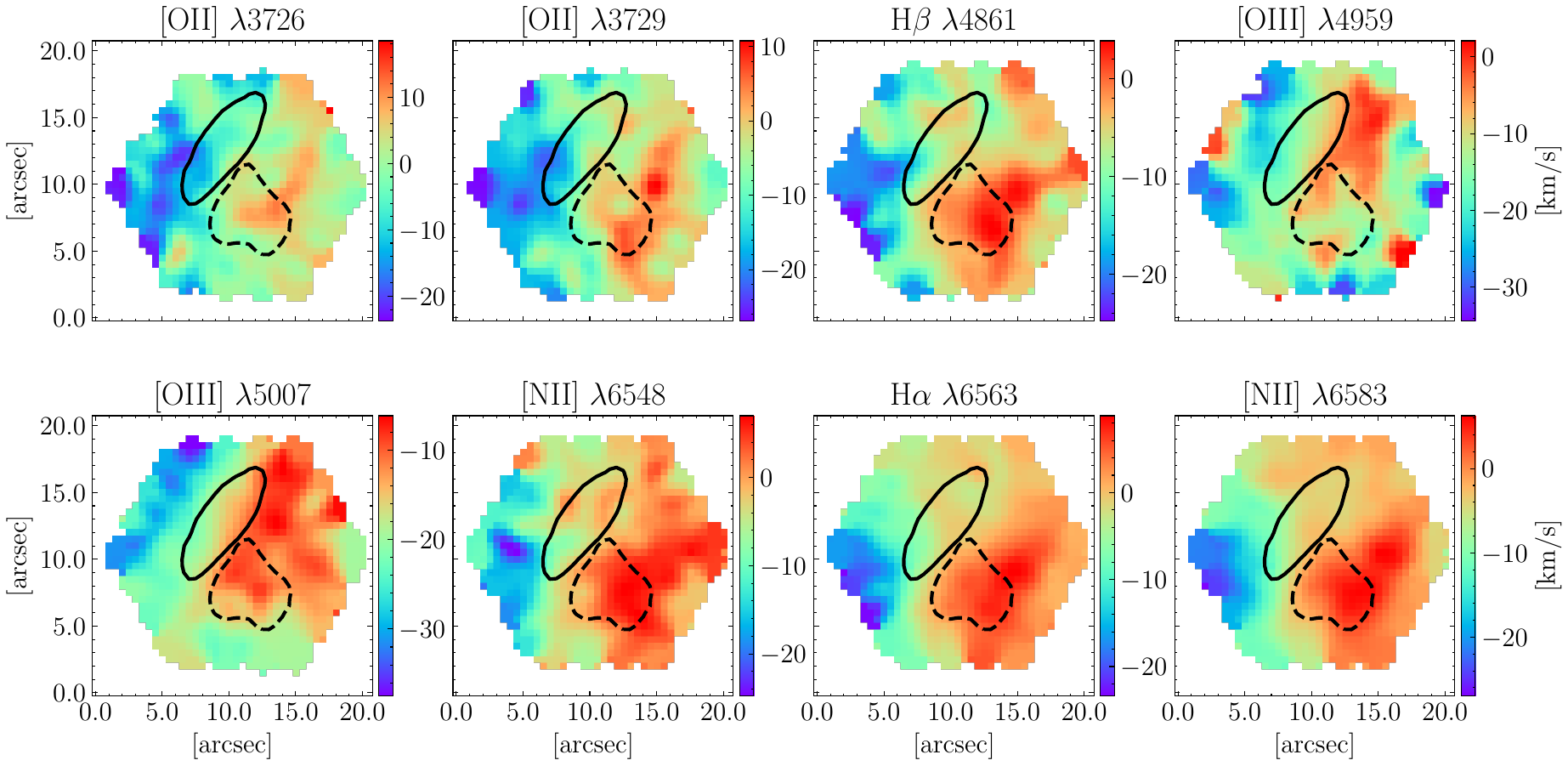}
	\caption{The velocity map of eight emission lines in the unit of km/s as measured from the MaNGA data. The bolded contours show 90\% quantiles of [O{\sc iii}] $\lambda$5007 (solid line) and H$\alpha$ $\lambda$6563 (dashed line) surface brightness.}
	\label{fig:velocity_filed_good}
\end{figure*}

\autoref{fig:surface_brightness_all} displays our measurements of 
the surface brightness map for all the 34 lines. In each panel we show 
the spaxels with S/N $>3$ only. 
According to the morphology of the maps, we find all the lines can be 
broadly divided into three groups. In the figure, the three groups of maps 
are shown separately with a blank gap inserted in between every two 
neighboring groups, while within each group the lines are ordered by
increasing wavelength. As can be seen, the maps in the first group 
present a remarkable lane of high brightness across the upper-left part 
of the MaNGA FoV. This group consists of seven high-ionization
lines and two [O{\sc ii}] auroral lines: [Ne{\sc iii}] $\lambda$3869, [O{\sc iii}] $\lambda$4363, [O{\sc iii}] $\lambda$4959, 
[O{\sc iii}] $\lambda$5007, [Ar{\sc iii}] $\lambda$7135, [O{\sc ii}] $\lambda$7320, 
[O{\sc ii}] $\lambda$7330,  [S{\sc iii}] $\lambda$9069 and [S{\sc iii}] $\lambda$9530. 
In what follows this group is referred to as ``Group I'', or 
``[O{\sc iii}] $\lambda$5007-like group'' to reflect the strongest fluxes of the 
[O{\sc iii}] $\lambda$5007 line in this group. We note that, for 
 [Ar{\sc iii}] $\lambda$7135 the lane feature is less pronounced than other lines
  of the same group, but we still include it in Group I considering that the
  emission in the lane region is the strongest and the only significant 
  feature of this line.

The second group is referred to as ``Group II'' or 
``H$\alpha$ $\lambda$6563-like group'', which is made up of 23
low-ionization or recombination lines:
H$\eta$ $\lambda$3836, H$\zeta$ $\lambda$3889, H$\epsilon$ $\lambda$3970, [S{\sc ii}] $\lambda$4069, [S{\sc ii}] $\lambda$4076, H$\delta$ $\lambda$4102, H$\gamma$ $\lambda$4340, H$\beta$ $\lambda$4861, [N{\sc i}] $\lambda$5200, [N{\sc ii}] $\lambda$5755, He{\sc i} $\lambda$5876, [O{\sc i}] $\lambda$6300, [O{\sc i}] $\lambda$6366, [N{\sc ii}] $\lambda$6548, H$\alpha$ $\lambda$6563, [N{\sc ii}] $\lambda$6583, [S{\sc ii}] $\lambda$6717, [S{\sc ii}] $\lambda$6731, [Ca{\sc ii}] $\lambda$7291, [Ca{\sc ii}] $\lambda$7324, [O{\sc ii}] $\lambda$7330, [C{\sc i}] $\lambda$8729, [C{\sc i}] $\lambda$9826 and [C{\sc i}] $\lambda$9852. The main feature of this group is a clump-like emission
region located in the lower-right part of the MaNGA FoV. Most of the 
maps in this group show a continuous area of clump, while some weak 
lines show more clumpy and less concentrated structures. We include both 
cases in this group, as long as the peak and the majority of the high
brightness area are located in the lower-right part of the MaNGA FoV. 
The third group includes only two lines, [O{\sc ii}] $\lambda$3726 and 
[O{\sc ii}] $\lambda$3729, which show clump-like features at both the 
position of the [O{\sc iii}] $\lambda$5007 lane and the position of the 
H$\alpha$ clump, with a bridge-like structure connecting the two clumps. 

To more clearly see the similarities of the maps within each group and 
the differences between different groups, in every panel we repeatedly 
show the brightness map of both [O{\sc iii}] $\lambda$5007
(the solid black contours) and H$\alpha$ $\lambda$6563
(the dashed contours). In the first group the lane-like feature shows 
up in all the maps with highly similar shapes and locations, while the 
clump-like feature seen in the H$\alpha$ map also similarly shows up in 
most of the maps in Group II. When comparing the two groups, it is noticeable 
that the lane-like feature in Group I has little spatial overlap with the 
clump-like feature in Group II. The spatial offset between the two groups 
of lines implies that the MaNGA region of the CCN is likely to be shocked 
by a blast wave moving from lower-right to upper-left. On the other hand,
however, the fact that the H$\alpha$ bright region is clump-shaped instead 
of lane-shaped cannot be simply explained by a single shock. 

\autoref{fig:velocity_filed_good} displays the velocity map for eight  
emission lines which have reliable velocity measurements, typically with
uncertainties less than $10\ \mathrm{km} / \mathrm{s}$. 
Overall, all the lines show relatively positive velocities in the lower-right 
region and negative velocities in the upper-left region. When 
comparing the velocity maps more closely, one can easily see 
both the high similarities within each group and the apparent differences
between different groups. For instance, the velocity maps of 
[O{\sc iii}] $\lambda$5007 and [O{\sc iii}] $\lambda$4959 show a lane-like structure 
in the same position as their surface brightness maps, a 
unique feature which is not seen in the velocity maps of other groups.
These velocity maps should in principle provide useful constraints 
to the physical processes behind the CCN. 
However, due to the limited spectral resolution of MaNGA, it is hard to 
decompose the different velocity components from the spectra. Therefore, 
we will use only the surface brightness measurements 
when constraining the shock models below in \autoref{sec:s_model_res}.

\section{Constraining shock models} \label{sec:s_model_res}

\subsection{The shock model}

In this section, we attempt to fit the MaNGA data by applying the shock 
model implemented in the {\tt MAPPINGS V} code \citep{Sutherland_2017, MAPPINGSV}.
The basic ideal of {\tt MAPPINGS V} is to calculate the flow solution of one-dimensional 
radiative shock models with parallel-slab geometry. The cooling function is 
established based on the {\tt CHIANTI 8} atomic database \citep{CHIANTI8},
and the pre-ionization effect is addressed via an iterative approach.
Other than the usually used Rankine-Hugoniot flow equations, in {\tt MAPPINGS V} 
a unique choosing of integral form of the conservation laws is employed to mitigate 
round-off errors. The reader is referred to \cite{Sutherland_2017} for more details 
about {\tt MAPPINGS V}.

\begin{figure}
	\centering
	\includegraphics[width=0.45\textwidth]{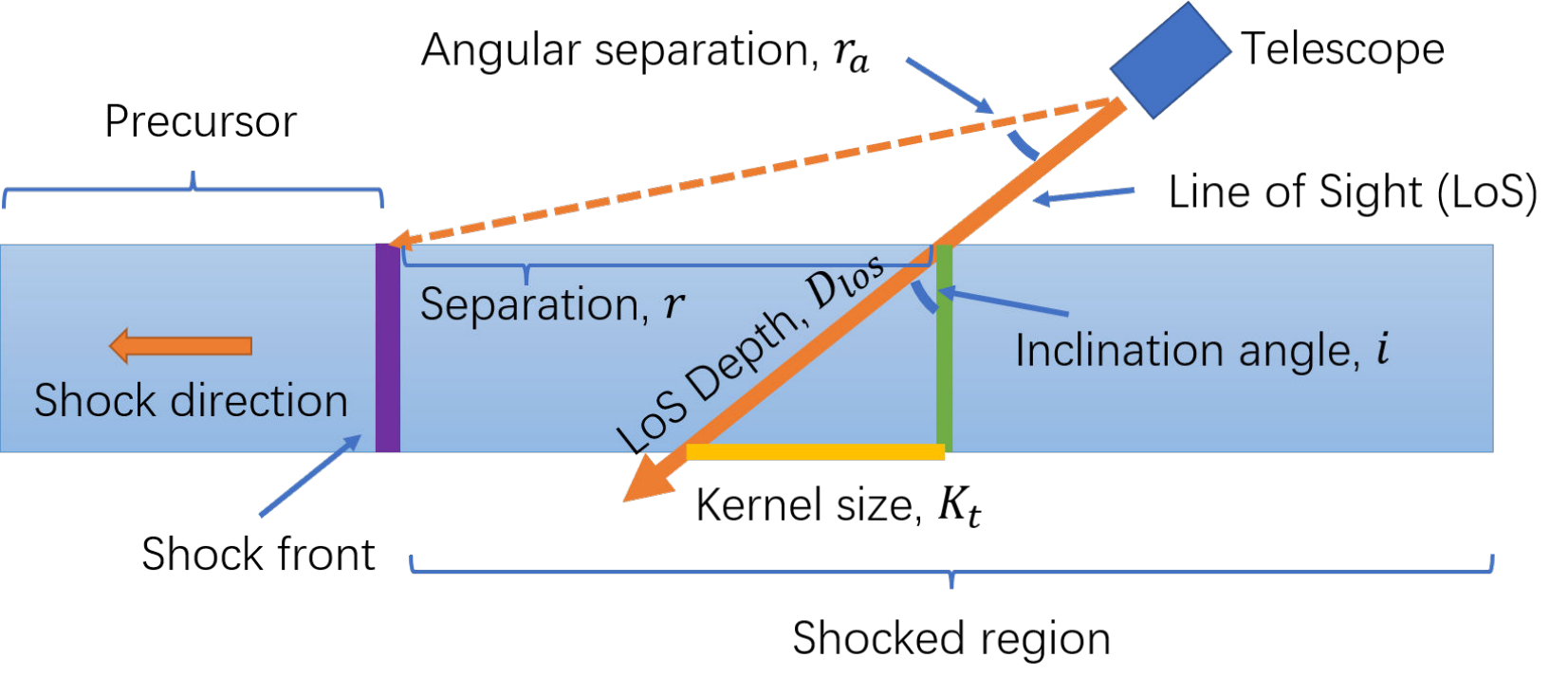}
	\caption{A sketch of the shock model adopted in our work. The 
		shock is assumed to have a parallel-slab geometry with a 
		depth of $D_{los}$ along the line of sight, and is orientated with an 
		inclination angle of $i$. The vertical purple 
		line marks the shock front, and the orange arrow in left side indicates the 
		shock direction.}
	\label{fig:geometry_model}
\end{figure}

To compare with the resolved MaNGA data, we use shock profiles calculated 
by the model, i.e.  the physical properties and emissivity as functions of the 
separation $r$ from the shock front, instead of the integrated spectrum 
which is the default output of the model. In this case, the shock front locates 
at $r = 0$, the region with $r > 0$ has already been shocked (thus referred to as 
``shocked region''), and the region with $r < 0$ is expected to be 
shocked in future (thus referred to as ``precursor''). The precursor may 
have been ionized by the photons escaping from the shocked region. 
Since {\tt MAPPINGS V} adopts parallel-slab geometry, in the direction 
perpendicular to the shock direction the physical properties and emissivity
are homogenous in the model. Therefore, the model is essentially a 
one-dimensional model, and we need to simplify the geometry of the 
observed region in consideration in order for appropriate comparisons 
between model and data. Our geometry model is illustrated in
\autoref{fig:geometry_model}. We assume the region is a ``slab'' of 
ionized gas with a certain thickness, oriented with an inclination angle 
of $i$ (the angle between the ``slab'' and the line of sight). 
As a result, the gas slab has a line-of-sight depth of $D_{los}$, and so 
the slab thickness is given by $D = D_{los}\cos(i)$.

For a given emission line, the observed surface brightness profile, 
$F_{\mathrm{obs}}(r_a)$ where $r_a$ is the angular separation
from the shock front, can then be modelled by
\begin{equation}\label{eqn:model_shock_profile}
    F_{\mathrm{obs}}(r_a) = G_{2.5} \star D_{los} [A(K_t) \star f(r_a)] + F_b.
\end{equation}
In the equation, the symbol ``$\star$'' represents the convolution operation. 
Here,  $f(r_a)$ is the intrinsic {\em emissivity} profile given by the model. 
For the calculation of $f(r_a)$, we adopt a distance of 150 pc to 
CCN in order to convert $r_a$ to physical separations
with the consideration of projection effect
(Uncertainties in CCN distance and the effects on our 
modelling will be discussed in \autoref{sec:distance}.). 
Due to the thickness and inclination of the gas region, the 
observed surface brightness at given $r_a$ is effectively an integral 
(along line of sight) of intrinsic emissivity at different 
{\em physical} separations. To take into account this effect, 
we first obtain an average surface brightness at $r_a$ by convolving 
$f(r_a)$ with the averaging filter $A(K_t)$ with a kernel size 
of $K_t=D_{los}\sin(i)$, which is then multiplied by the line-of-sight 
depth $D_{los}$ to give the total surface brightness, as can 
be seen from the first term at the right hand of \autoref{eqn:model_shock_profile}.
This term is further convolved by $G_{2.5}$, 
a Gaussian filter with $\mathrm{FWHM} = 2.5''$, which
models the point spread function (PSF) of the MaNGA data.
Finally, the last term of the equation represents an additional 
component, $F_b$,  to account for the background emission 
which could be either a constant background/foreground 
in the line of sight (signal or noise, or both), or produced by 
ionizing mechanisms other than the shock.
As we will show in \autoref{sec:model_const}, the model with a constant background cannot explain our data. Thus, we have an extended model in \autoref{sec:model_gaussian}.

In addition, we also need to determine the location of the shock front
and the {\em projected} direction of the shock. For simplicity, 
we fix the shock direction as the diagonal line going 
from the lower-right corner to the upper-left corner in the 
MaNGA FoV. This direction is perpendicular to the lane-like structure 
in the [OIII] surface brightness map, at which we have also seen 
sharp jumps in some line ratios 
(see \autoref{fig:clean_region_on_line_ratio}; the shock 
direction is indicated by the arrow in each panel). The shock should 
have moved from the H$\alpha$ clump region to the [O{\sc iii}] 
lane, so that the latter should be closer to the shock front.
The location of the shock front cannot be determined by 
simple considerations, and this is modeled by a free parameter,
$\Delta S$, the angular distance of the shock front to the 
central spaxel of the MaNGA FoV.

The physical parameters of the shock considered in this work 
include the pre-shock hydrogen density $n_\mathrm{pre}$, the 
shock velocity $v_\mathrm{sh}$, and the transverse magnetic field $B$. 
Considering the fact that CCN is projected in and also possibly 
located in the Orion-Eridanus superbubble, we adopt the metal abundance of the Orion nebula provided in {\tt MAPPINGS V}. We note that changes 
in abundance can result in complex and non-linear 
modifications of shock properties. For instance, under certain parameters, 
a shock with a high metallicity can produce a higher 
[O{\sc iii}] $\lambda$5007/H$\beta$ $\lambda$4861 ratio when 
compared to its low metallicity counterpart. However, we find that,
if the metallicity and abundance are not fixed, the model will have 
too many degrees of freedom and will be very hard to be constrained 
well with the available data. Therefore, we opt to fix the metallicity and 
abundance in this work and leave the effect of potential variations 
in metallicity or abundance to be studied in future when more data 
becomes available.

\subsection{Model 1:  shock with a constant background} \label{sec:model_const}

\begin{figure*}
	\centering
	\includegraphics[width=0.7\textwidth]{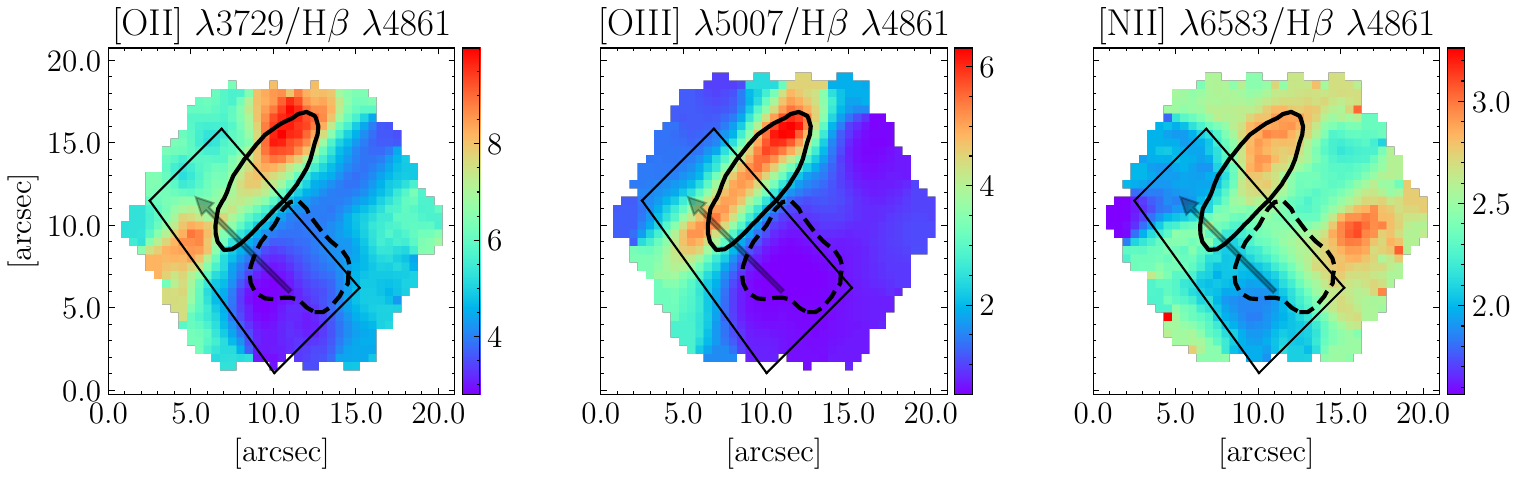}
	\caption{Maps of three line ratios: [O{\sc ii}] $\lambda$3729/H$\beta$ $\lambda$4861, 
		[O{\sc iii}] $\lambda$5007/H$\beta$ $\lambda$4861 and [N{\sc ii}] $\lambda$6583/H$\beta$ $\lambda$4861.
		The ``clean region'' used to constrain the shock in Model 1 (see \autoref{sec:model_const}) 
		is indicated by a black quadrilateral. The gray arrow indicates the 
		propagating direction of the shock. The bolded contours show 90\% quantiles of [O{\sc iii}] $\lambda$5007 (solid line) and H$\alpha$ $\lambda$6563 (dashed line) surface brightness.}
	\label{fig:clean_region_on_line_ratio}
\end{figure*}

\begin{figure*}
	\centering
	\includegraphics[width=0.85\textwidth]{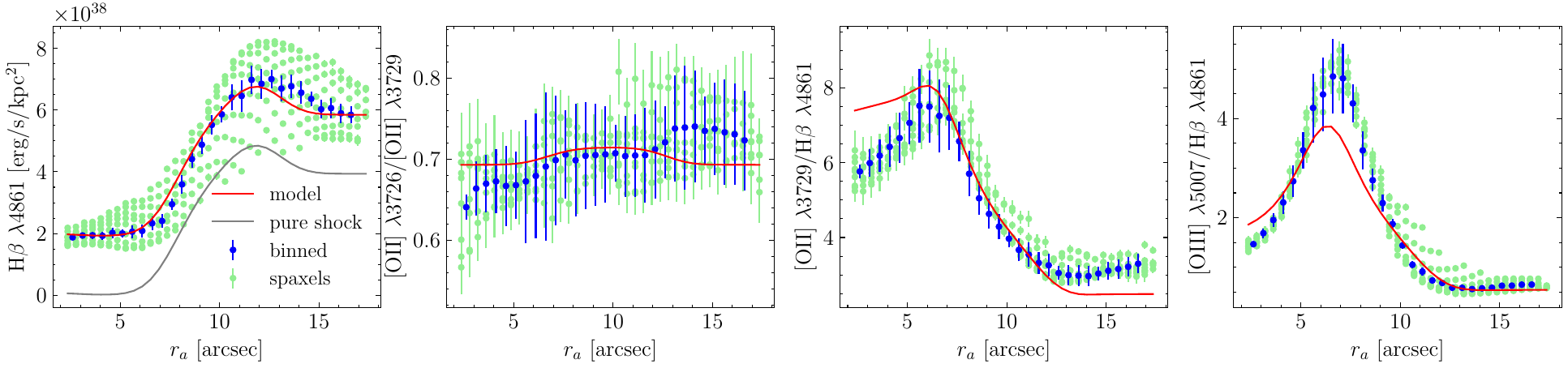}
	\caption{The shock profiles of H$\beta$ $\lambda$4861 surface brightness as well as emission lines 
		ratios [O{\sc ii}] 3726/3729, [O{\sc ii}] $\lambda$3729/H$\beta$ $\lambda$4861 and [O{\sc iii}] $\lambda$5007/H$\beta$ 4861, 
		as measured from the ``clean region'' in the MaNGA data and predicted by the best-fit Model 1. 
		The green dots show the measurements for individual spaxels and the blue dots show the mean 
		at given $r_a$ (angular distance from shock front). The solid red line is the best-fit model, while 
		the grey line in the leftmost panel shows the model prediction without the constant background.}
	\label{fig:shock_model_1d}
\end{figure*}

\begin{deluxetable*}{cccccccc}
	% \tablenum{2}
	\tablecaption{The best-fit parameters of shock models with 1 $\sigma$ uncertainty \label{tab:model_parameters}}
	\tablewidth{0pt}
	\tablehead{
		\colhead{Model} & \colhead{$B$} & \colhead{$n_\mathrm{pre}$} & \colhead{$v_\mathrm{sh}$} & \colhead{$K_t$} & \colhead{$D_{los}$} & \colhead{$\Delta S$} & \colhead{$i = \arcsin (K_t/D_{los})$}\\
		\colhead{} & \colhead{$[\mathrm{\mu G}]$} & \colhead{$[\mathrm{cm}^{-3}]$} & \colhead{$[\mathrm{km}/\mathrm{s}]$} & \colhead{$[\mathrm{AU}]$} & \colhead{$[\mathrm{AU}]$} & \colhead{$[\mathrm{arcsec}]$} & \colhead{$[\mathrm{deg}]$}
	}
	% \decimalcolnumbers
	\startdata
	Model 1 (H$\beta$) & $1.4 \pm 1.2 \times 10^{-4}$ & $1.3\pm 0.2$ & $112\pm 7$ & $687\pm 65$ & $1990\pm 453$ & $7.2 \pm 1.4$ & $22\pm7$ \\
	Model 1 (H$\alpha$) & $0.1 \pm 0.1 \times 10^{-5}$ & $1.3\pm 0.2$ & $110\pm 9$ & $595\pm 60$ & $1890\pm 415$ & $7.4 \pm 2.9$ & $19\pm5$ \\
	Model 2 & $1.6 \pm 0.3$ & $1.4\pm 0.2$ & $133\pm 5$ & $419\pm 324$ & $1150\pm 328$ & $17.2 \pm 3.8$ & $19\pm14$ \\
	% $\mathrm{[OII]}$ $\lambda$3729 & 3730.1 & 0.52 & Possible over-subtracted line\\
	% H$\zeta$ $\lambda$3889 & 3890.1 & 0.40 & Too strong to be real\\
	% H$\beta$ $\lambda$4861 & 4863.1 & 0.44 & Can originate from geocoronal emission line\\
	% H$\alpha$ $\lambda$6563 & 6564.7 & 2.20 & Can originate from geocoronal emission line\\
	% $\mathrm{[SII]}$ $\lambda$6717 & 6718.8 & 0.43 & Possible over-subtracted line\\
	% $\mathrm{[SII]}$ $\lambda$6731 & 6733.2 & 0.27 & Possible over-subtracted line\\
	\enddata
\end{deluxetable*}

\begin{figure}
	\centering
	\includegraphics[width=0.45\textwidth]{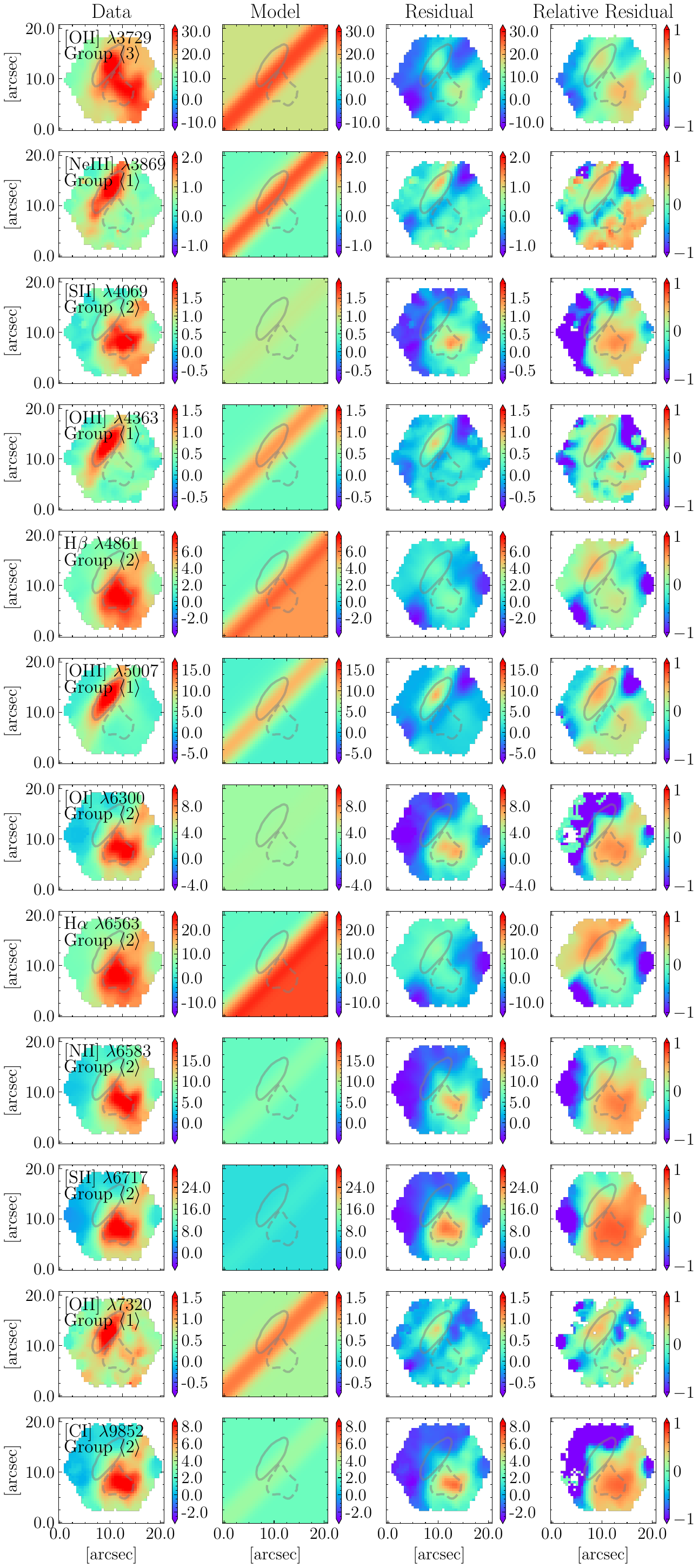}
	\caption{Surface density maps of 12 emission lines (panels from top to bottom).
		For each line, panels from left to right show the observed map, 
		the predicted map by Model 1, and the absolute and 
		relative residual maps. All the maps are color-coded by surface 
		brightness in units of $10^{38}$ erg/s/kpc$^2$. The gray contours show 90\% quantiles of [O{\sc iii}] $\lambda$5007 (solid line) and H$\alpha$ $\lambda$6563 (dashed line) surface brightness.}
	\label{fig:shock_model_1d_residual}
\end{figure}

\begin{figure}
	\centering
	\includegraphics[width=0.45\textwidth]{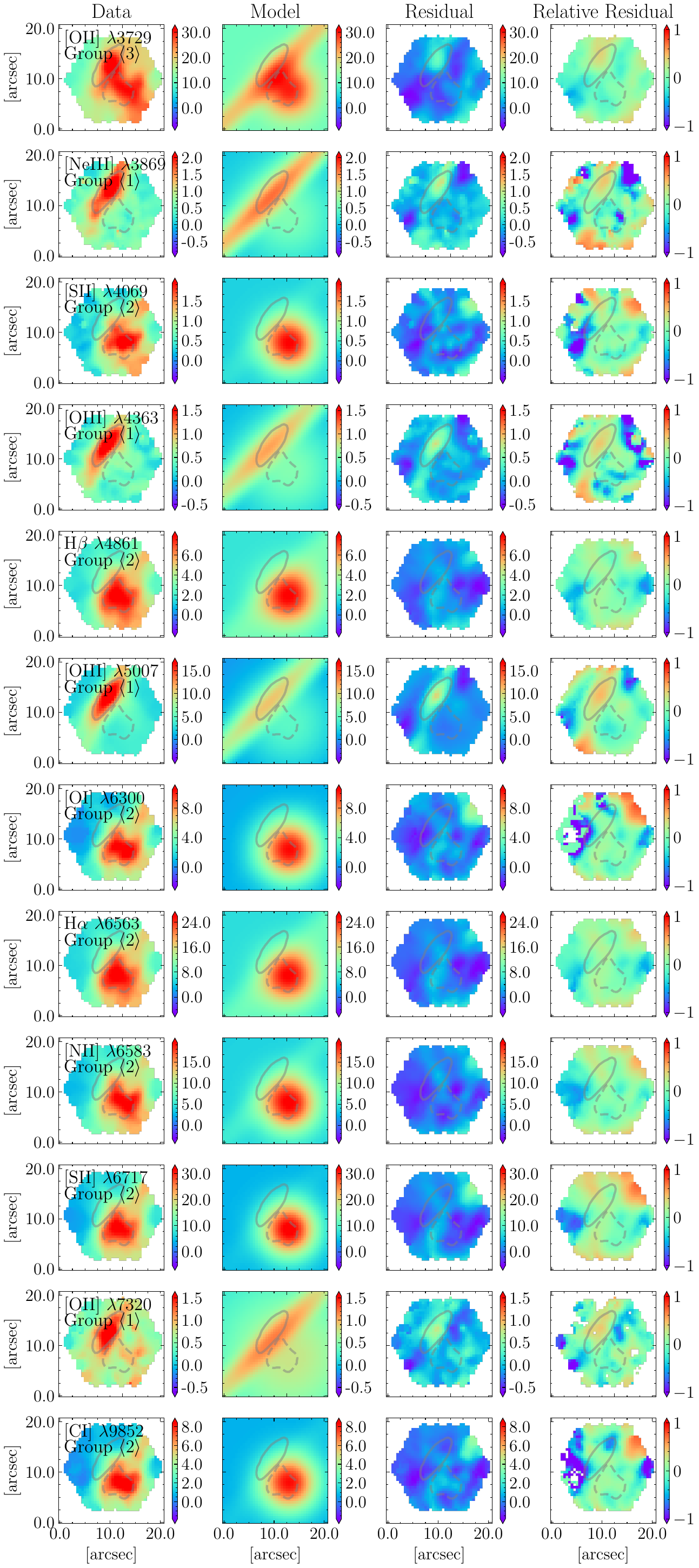}
	\caption{Same as Figure \ref{fig:shock_model_1d_residual} but for Model 2.}
	\label{fig:shock_model_g+s_residual}
\end{figure}

We first consider the model with a constant background component.  
We have attempted to fit the shock and the background components 
simultaneously using all available line and line ratio measurements
from the full MaNGA FoV, but finding ourselves unable to obtain a 
reasonably well constrained solution. We thus opt for a two-step approach,
in which  we first constrain the shock parameters, including both the 
physical parameters ($n_\mathrm{pre}$, $v_\mathrm{sh}$, $B$) and 
the geometric parameters ($K_t$, $D$, $\Delta S$), by applying the model 
to a selected set of emission lines and line ratios from a ``clean region'' 
selected out of the MaNGA FoV, and then we constrain the background 
component $F_b$ for all the emission lines from the full MaNGA FoV 
but fixing the parameters of the shock obtained from the first step.  

The measurements used in the first step include 
the surface brightness profile of H$\beta$ $\lambda$4861 
representing the H$\alpha$ $\lambda$6563-like group, the profile of
[O{\sc iii}] $\lambda$5007 representing the 
[O{\sc iii}] $\lambda5007$-like group, and the profiles of 
[O{\sc ii}] $\lambda$3729 and
[O{\sc ii}] $\lambda$3726 from the third group.
We include both lines from the [O{\sc ii}] doublet because they are 
comparably strong, unlike the other two groups in which one or two lines 
are much stronger than the other lines. For the H$\alpha$ $\lambda$6563-like group,
we use H$\beta$ rather than H$\alpha$ because the former
is closer to [O{\sc iii}] $\lambda5007$ in wavelength, and so the effect
of dust extinction can be mitigated (although the dust extinction 
in CCN should be very weak; see \autoref{sec:dust_extinction} for discussion). 
In fact, we have repeated 
the fitting procedure using H$\alpha$ to replace H$\beta$ in 
the first step, and we find very similar results (see \autoref{tab:model_parameters}). 

The ``clean region'' used in the first step is indicated by a black quadrangle
in each panel of \autoref{fig:clean_region_on_line_ratio}. This region is 
selected to have simple shock profiles in the line ratios along the 
projected direction of the shock (as indicated by the arrow).
The exact boundary of this ``clean region'' is manually specified to maximize 
the number of spaxels included in the modelling, and at the same time to 
minimize the contamination of more complex features.  We notice that the 
apparent difference between regions inside and outside the ``clean region'' 
is a hint that the whole region cannot be simply modeled with a single shock.
We will come back and discuss this point in \autoref{sec:num_direc_shock}. 

% [OII] $\lambda \lambda$3726, 3729 doublet because they are very strong, have a unique surface brightness distribution, and can be well-fitted with [OIII] $\lambda$5007 and H$\beta$ $\lambda$4861 at the same time. Other strong lines such as H$\alpha$ $\lambda$6563, [NII] $\lambda$6583, [SII] $\lambda$6717 are not included because the model can not converge if we include these lines (the model 2 can fit these lines well, and we will discuss that model in Section \ref{sec:model_gaussian}). We note that the choice of the lines used in the first step is a bit arbitrary. We have tried to substitute H$\beta$ $\lambda$4861 with H$\alpha$ $\lambda$6563 and carry out the corresponding fitting procedure. Then, we found this alternative selection generates similar results of physical and geometrical parameters (See Table \ref{tab:model_parameters}). The over-subtracted lines are not added back, because this effect can be absorbed into constant background components. We do not apply the S/N cut in our fitting but just directly use all available spaxels. Because in our case the S/N is dominated by the strength of the signal. Thus, the low S/N contains the information that the emission in this spaxel is not significant, which should be reproduced by the model. Meanwhile, we use the inverse variance of the line ratio as the weight of each spaxel, hence our model will not be ``misled'' by these low S/N spaxels.

It is time consuming to calculate the shock properties for a given set of model
parameters, and so the calculation would be very expensive when 
the geometry parameters are fitted simultaneously with the shock parameters.
In practice, we use a ``zoom-in'' grid fitting approach.  We start with a 
coarse grid of the shock model parameters, and find the best-fit geometry 
parameters for each node of the grid by minimizing the reduced $\chi^2$.
The best-fit model in this grid is then given by the node at which the
reduced $\chi^2$ is the smallest over the whole grid,
 $\chi^2_{\mathrm{min}}$.  A weight is then given to each of the nodes 
 by $W = \mathrm{exp} (-\frac{1}{2} (\chi^2 - \chi^2_{\mathrm{min}}))$, 
 and the 1 $\sigma$ uncertainty of each parameter is calculated from 
the weighted average of all the nodes. If the uncertainties are significantly 
smaller than the step sizes of the grid, we construct a finer grid around the 
best-fit node and repeat the above process until the uncertainties 
are comparable with the step sizes of the grid. 

The best-fit results from the first step are shown in \autoref{fig:shock_model_1d}, 
where we plot the observed profiles from the clean region and the best-fit shock profiles. 
The best-fit model parameters and uncertainties are listed in \autoref{tab:model_parameters}.
The model parameters constrained by using the H$\alpha$ line instead of H$\beta$ 
are also listed in the table. Generally, all the profiles are well-fitted in the whole range of 
$r_a$ probed, except that the model slightly overpredicts (underpredicts) 
the ratio of [O{\sc ii}] $\lambda$3729/H$\beta$ $\lambda$4861 at the smallest (largest) 
$r_a$ (see the third panel of the figure) and the peak of [O{\sc iii}] $\lambda$5007/H$\beta$ $\lambda$4861 is a bit shallower in model. It is encouraging that the models with 
H$\beta$ and H$\alpha$ give rise to very similar constraints for all the model parameters.
Accordingly, the shock has a velocity of $v_\mathrm{sh}\sim110$ km/s, a pre-shock hydrogen
density of $n_\mathrm{pre}\sim1.3$ cm$^{-3}$, and a transverse magnetic field 
which is very close to zero (though with relatively large uncertainties). 
The shock velocity is larger than the value found by Z97, which is $40\ \mathrm{km}/\mathrm{s}$. 
This discrepancy may not be surprising, considering that the targeted position
of the spectroscopic study in Z97 is outside the MaNGA FoV. 
The geometry parameters of our best-fit model show that the direction of 
shock is almost perpendicular to our line of sight, with an inclination angle of 22
deg (with H$\beta$) or 19 deg (with H$\alpha$). The shock front is about 7 arcsec 
from the central spaxel, thus close to the upper-left edge of the MaNGA FoV.

\autoref{fig:shock_model_1d_residual} shows the best-fit results for some 
lines as obtained from the second step, in which we fix the shock model 
parameters and further constrain the background component for individual lines. 
%The fact that some lines we measured by a single Gaussian function are actually more than one of lines blended together (for example, the H$\epsilon$ $\lambda$3970 is blended with [NeIII] $\lambda$3967) is considered in our fitting. The fitting results for some lines are shown in Figure \ref{fig:shock_model_1d_residual} as an example, 
For a given line, the figure displays the maps of the observed surface brightness, 
the surface brightness predicted by the model, the residual defined as 
the difference between the observation and model prediction ($\mathrm{Res}$), 
and the relative residual ($\mathrm{RR}$) defined as the ratio of the residual 
relative to the geometric mean of the surface brightness measurement and 
its error for each spaxel, i.e. 
$\mathrm{RR} = \mathrm{Res}/\sqrt{S^2 + \mathrm{err}^2}$, 
where $S$ is the surface brightness of the emission line in the given spaxel, 
and $\mathrm{err}$ is the 1$\sigma$ uncertainty of the surface brightness.
For spaxels with high S/N, the relative residual is effectively equal to fractional error, 
while for low S/N spaxels, the relative residual is reduced to the ratio between 
residual and measurement error, thus telling whether the residual can be 
explained by data uncertainties.

As can be seen, our model successfully captures some main features in the data, 
such as the [O{\sc iii}] lane-like structure and the overall difference in surface 
brightness between the upper-left and lower-right regions. However, the 
model substantially underpredicts or overpredicts the surface brightness 
for certain regions and for all the lines considered, which can be easily 
identified from the relative residual maps. 

\subsection{Model 2: shock with a 2D Gaussian component} \label{sec:model_gaussian}

The above result strongly suggests that the $F_b$ component in our model 
cannot be simply a constant background. We have considered a different model 
in which we add another shock, that is, the observed 
region is jointly ionized by two distinct shocks, but found such kind of model can not 
fit the data well (see \autoref{sec:discuss_gaussian}).

In this section we present our second model (Model 2 hereafter) in which 
we include a two-dimensional (2D) Gaussian component in addition to the 
shock component and the constant background. In this case, \autoref{eqn:model_shock_profile} is rewritten as
\begin{equation}\label{eqn:model_shock_profile2}
	F_{\mathrm{obs}}(x, y)=G_{2.5}\star D\{A(K_{t})\star f[r_{a}(x, y)]\}+F_{G}(x, y)+F_{b},
\end{equation}
where the new term $F_{G}(x,y)$ is an isotropic 2D Gaussian function
($\sigma_x = \sigma_y$). Note that 
the observed surface brightness and the intrinsic emissivity are also written 
as two-dimensional functions of spaxel position $(x,y)$. This is because 
the total surface brightness in both data and model can no longer be 
described by a one-dimensional profile after the Gaussian component is included. 
Different emission lines share the same Gaussian center ($x, y$), but have their 
own width $\sigma_x$ and amplitude $A$. 
The fitting procedure is divided into two successive steps 
as for the first model (Model 1 hereafter). Different from Model 1,
for the first step of Model 2 we use 
more emission lines and all available spaxels from the MaNGA FoV to 
constrain the shock parameters, considering the larger number of model 
parameters and the more complex model components. Specifically, 
we use the surface brightness profiles of [O{\sc ii}] $\lambda$3729, 
H$\beta$ $\lambda$4861, [O{\sc iii}] $\lambda$5007, 
H$\alpha$ $\lambda$6563, [N{\sc ii}] $\lambda$6583 and [S{\sc ii}] $\lambda$6717. 

This model can fit the data much better than Model 1, although the physical 
origin of the Gaussian component still remains unclear. 
The best-fit parameters and uncertainties of the shock component 
are listed in Table \ref{tab:model_parameters}. Compared with Model 1, 
the shock in Model 2 has a slightly higher pre-shock hydrogen density (1.4 $\mathrm{cm}^{-3}$ versus 1.3 $\mathrm{cm}^{-3}$), slightly larger 
velocity (133 km/s versus $\sim$110 km/s), and a much stronger magnetic field
 (1.6 $\mu$G versus $\lesssim10^{-4}$ $\mu$G). For the geometric parameters, 
both $K_t$ and $D$ are slightly smaller than those from Model 1, while $\Delta S$ increases from $\sim$7 arcsec to 17.2 arcsec. 
The inclination angles from the two models are quite similar, at around 19 deg.

The fitting results from the second step are shown in 
\autoref{fig:shock_model_g+s_residual}, for the same set of emission lines as 
in the previous figure. As can be seen, the $\mathrm{RR}$ is significantly 
reduced in most spaxels and for all the lines when compared to the fitting 
results of Model 1. 
From the model-predicted maps, one can easily find both the differences 
between different groups of lines and the similarities within the same group: 
the three lines from the [O{\sc iii}] $\lambda5007$-like group are uniformly 
dominated by the lane-like structure, the Gaussian component dominates 
all the lines from the H$\alpha$ $\lambda$6563-like group, and both components are 
presented in the map of [O{\sc ii}] $\lambda3729$ with comparable amplitudes.
We have examined the ratios of different lines for the Gaussian 
component alone, finding them to be comparable to the expected line ratios of a clump of gas ionized by the photons from the shocked region (\autoref{sec:discuss_gaussian}). Although the fits are substantially improved 
with respect to Model 1, one can still see patterns of significant residuals.
For instance, the maps tend to present positive values of $\mathrm{RR}$
in the left corner for the [O{\sc iii}] $\lambda$5007-like lines as well as 
the upper-right edge for the H$\alpha$ $\lambda$6563-like lines. At the border of the 
MaNGA FoV, these regions could be adjacent to other regions 
of the CCN that may be ionized in a different way. In addition, a 
lane-like residual pattern can be seen in/around the original [O{\sc iii}] lane 
structure, for all the emission lines but with different orientations for 
different line groups. This pattern is likely to be induced by another 
shock along the upper-right to lower-left direction. We will come back 
and discuss on this point in \autoref{sec:num_direc_shock}.

\begin{figure}
	\centering
	\includegraphics[width=0.4\textwidth]{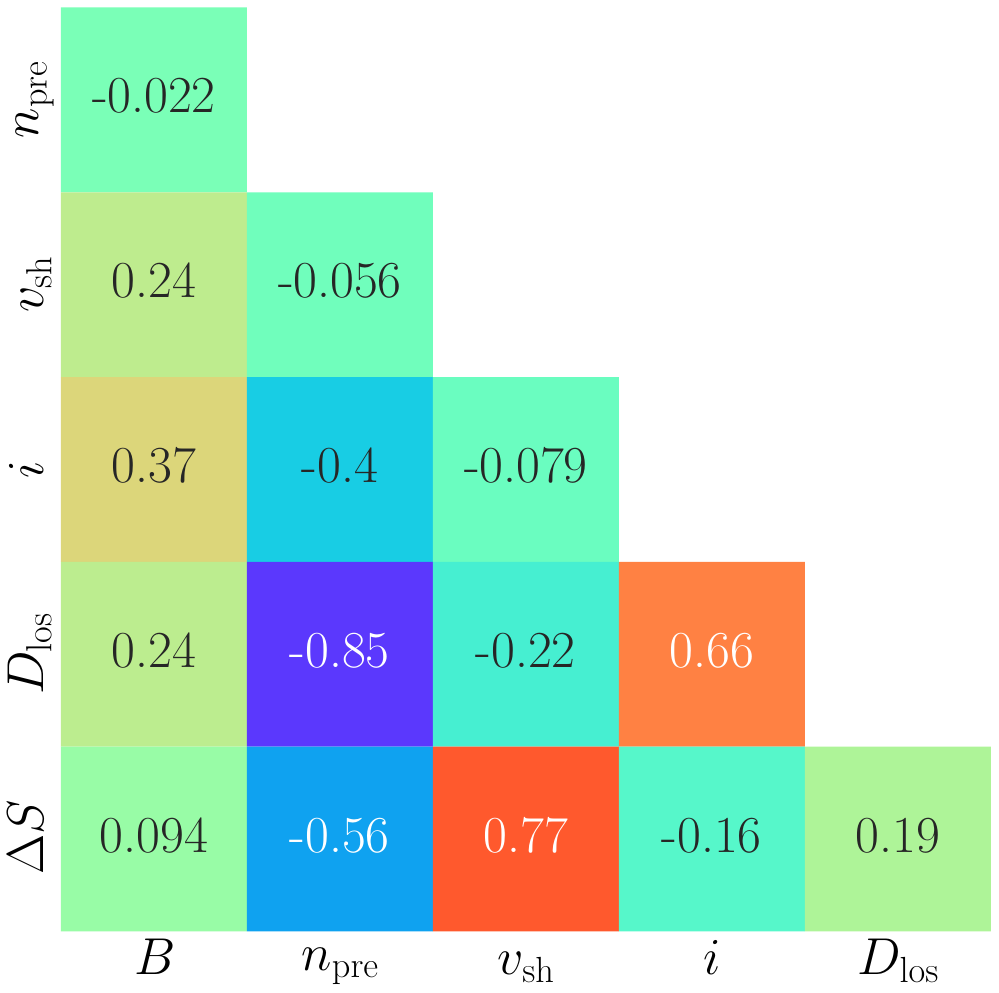}
	\caption{The weighted Pearson correlation coefficients for best-fit parameters of Model 2.}
	% The value of best fitting model and corresponding uncertainty for each parameter is shown in the diagonal.
	\label{fig:parameter_correlation}
\end{figure}

We evaluate the correlations between model parameters 
by calculating the weighted Pearson correlation coefficient 
$r_\mathrm{pearson}$ based on the grid generated during the fitting process.
The result is shown in \autoref{fig:parameter_correlation} for 
the six free parameters describing the shock and the geometry.
The shock parameters ($v_\mathrm{sh}$, $n_\mathrm{pre}$, $B$) show weak correlations between each other, but they all show
	strong correlations with one or more geometric parameters.
This analysis suggests that, though not very strong, our  
model suffers from degeneracies caused by the projection
effect and the uncertainties in the geometry of the gas region. 
We would like to point out that the degeneracies between model 
parameters are taken into account in the 1$\sigma$ uncertainties 
of our best-fit model parameters, which are actually the weighted 
standard deviation of the marginalized distributions. 

\begin{figure*}
	\centering
	\includegraphics[width=0.95\textwidth]{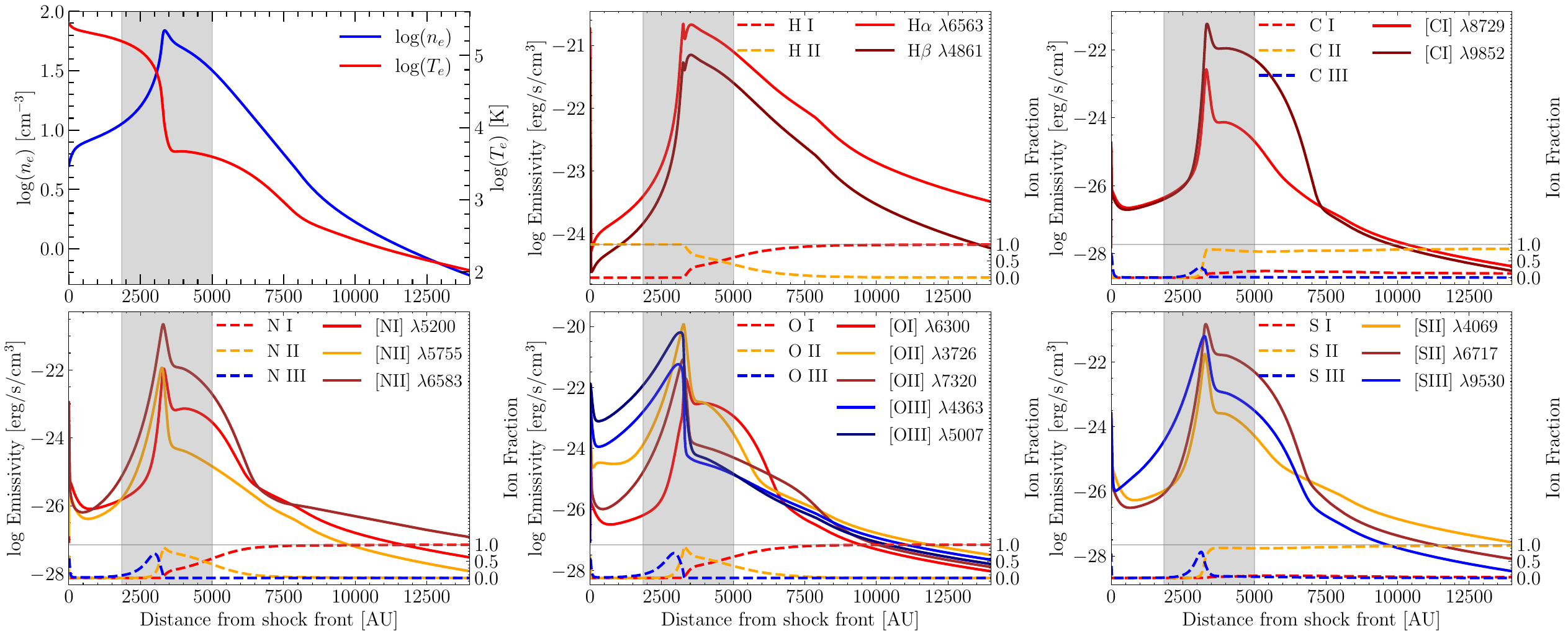}
	\caption{In the top left panel, we present the intrinsic shock profiles of the electronic density ($n_e$) in blue and the temperature ($T_e$) in red, as predicted by Model 2. The remaining panels display the ion fractions of hydrogen (H), carbon (C), nitrogen (N), oxygen (O), and sulfur (S), as well as the emissivity of the corresponding emission lines. These panels are arranged from left to right and top to bottom. The shaded region in each panel represents the coverage of the MaNGA FoV.}
	\label{fig:shock_model_g+s_ne_Te}
\end{figure*}

\begin{figure*}
	\centering
	\includegraphics[width=0.9\textwidth]{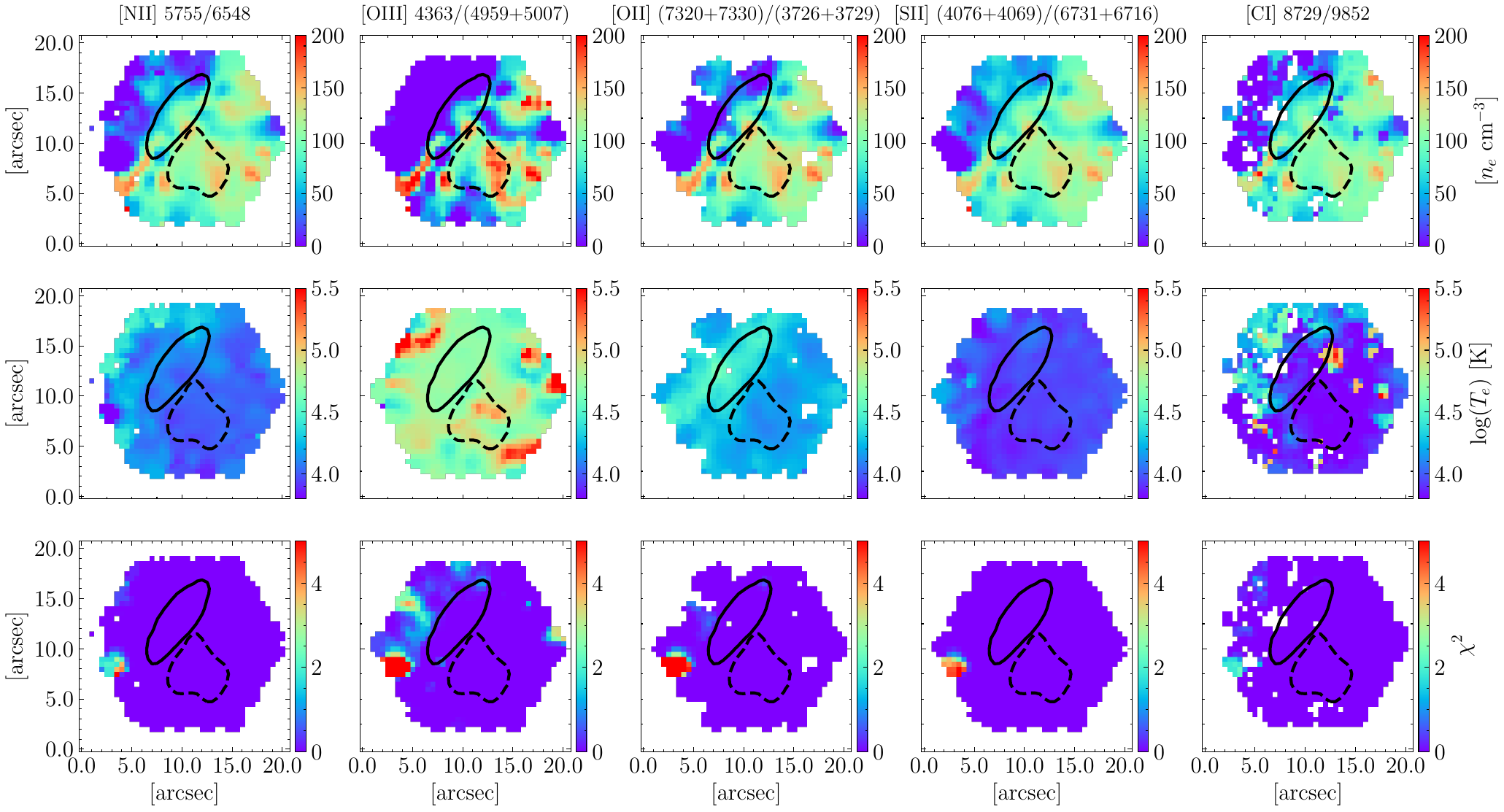}
	\caption{The electron density map (first row) and temperature map (second row) 
		as measured from the combinations of density-sensitive line ratio 
		[S{\sc ii}] 6731/6717 and different temperature-sensitive line ratios
		(panels from left to right): 
		[N{\sc ii}] 5755/6548, [O{\sc iii}] 4363/(4959+5007), [O{\sc ii}] (7320+7330)/(3726+3729), [S{\sc ii}] (4076+4069)/(6731+6716), and [C{\sc i}] 8729/9852. 
		The bottom panels show the corresponding $\chi^2$ maps. The bolded contours in each panel show 90\% quantiles of [O{\sc iii}] $\lambda$5007 (solid line) and H$\alpha$ $\lambda$6563 (dashed line) surface brightness.}
	\label{fig:nT_map}
\end{figure*}

\subsection{Shock profiles of electron temperature and density} \label{sec:Tene_model}

The top-left panel of \autoref{fig:shock_model_g+s_ne_Te} displays the profiles of both electron 
temperature $T_e$ (the red curve) and density $n_e$ (the blue curve) 
in the shock component as predicted by our best-fit model (Model 2).
Plotted in the other panels are the emissivity profiles (solid lines) and the 
	corresponding ion fraction profiles (dashed lines) of hydrogen (H), carbon (C), nitrogen (N), oxygen (O) and sulfur (S). 
As can be seen, the MaNGA FoV (indicated by the gray band) happens to capture the jump in density and 
temperature (and subsequently, the line emissivities) at $r \sim 3500$ AU.

An alternative and more commonly-adopted way to 
measure temperatures and densities of ionized gas is to use a set of selected 
emission line ratios that are sensitive to $n_e$ or $T_e$ as diagnostics,
based on a given atomic database. 
It is interesting to compare the $T_e$ and $n_e$ profiles predicted by our 
model with those calculated directly from emission line ratios.
We use {\tt PyNeb} \citep{pyneb}
to calculate $T_e$ and $n_e$ for each spaxel within the MaNGA FoV
based on our measurements of emission line ratios. For consistency, 
we use {\tt CHIANTI 8} \citep{CHIANTI8}, the same atomic database as 
used in {\tt MAPPINGS V}. In fact, we have explored alternative databases 
available in {\tt PyNeb}, finding them to provide similar $n_e$ and $T_e$.
We use the line ratio of [S{\sc ii}] 6731/6717 as the diagnostic of $n_e$, 
and use five different line ratios as diagnostics of $T_e$: [N{\sc ii}] 5755/6548, 
[O{\sc iii}] 4363/(4959+5007), [O{\sc ii}] (7320+7330)/(3726+3729), [S{\sc ii}] (4076+4069)/(6731+6716), 
and [C{\sc i}] 8729/9852. 
For a given pair of $T_e$-sensitive and $n_e$-sensitive
diagnostics, we obtain the measurements of $T_e$ and $n_e$ simultaneously 
for each spaxel in the MaNGA FoV. Therefore, five combinations of $T_e$-sensitive line ratios and 
the $n_e$-sensitive 
line ratio give rise to a set of five measurements of the $T_e$ and $n_e$ 
maps, which are shown in the top and middle rows in \autoref{fig:nT_map}. 

Each combination utilizes different $T_e$-sensitive line ratios while keeping the $n_e$-sensitive line ratio fixed as [S{\sc ii}] 6731/6717.
The alternative $n_e$-sensitive line ratio [O{\sc ii}] 3726/3729 is not utilized due to its significant uncertainty, as illustrated in the second panel of \autoref{fig:shock_model_1d}. This uncertainty obscures the true density structure and generates a misleading clumpy artificial density distribution. The primary cause of this significant uncertainty stems from that the [O{\sc ii}] $\lambda\lambda$ 3726,3729 doublets is not well resolved in the MaNGA spectra, as can be seen in the first two panels of \autoref{fig:lines_fitting_21_21}. Therefore, using [O{\sc ii}] 3726/3729 would risk over-interpreting the data. However, it would be worthwhile to investigate the consistency between the density measured by [O{\sc ii}] 3726/3729 and [S{\sc ii}] 6731/6717 when higher quality data becomes available.
It is important to note that, although $n_e$ is mainly constrained by [S{\sc ii}] 6731/6717, the different $T_e$-sensitive lines lead to slight differences in the derived $n_e$. This is why the $n_e$ maps in the first row of \autoref{fig:nT_map} differ quantitatively, although they qualitatively show good agreement.
The bottom row in the same figure shows the corresponding $\chi^2$ 
maps. As can be seen, 
the $\chi^2$ values are close to zero in most spaxels of a given map, 
indicating that all the lines relevant for measuring $T_e$ and $n_e$ 
are well-fitted. We note that, we have added back the over-subtracted sky flux 
when calculating [SII] $\lambda\lambda$ 6717,6731, although this 
correction has no significant effect on our results.

\begin{figure*}[ht!]
	\centering
	\includegraphics[width=0.95\textwidth]{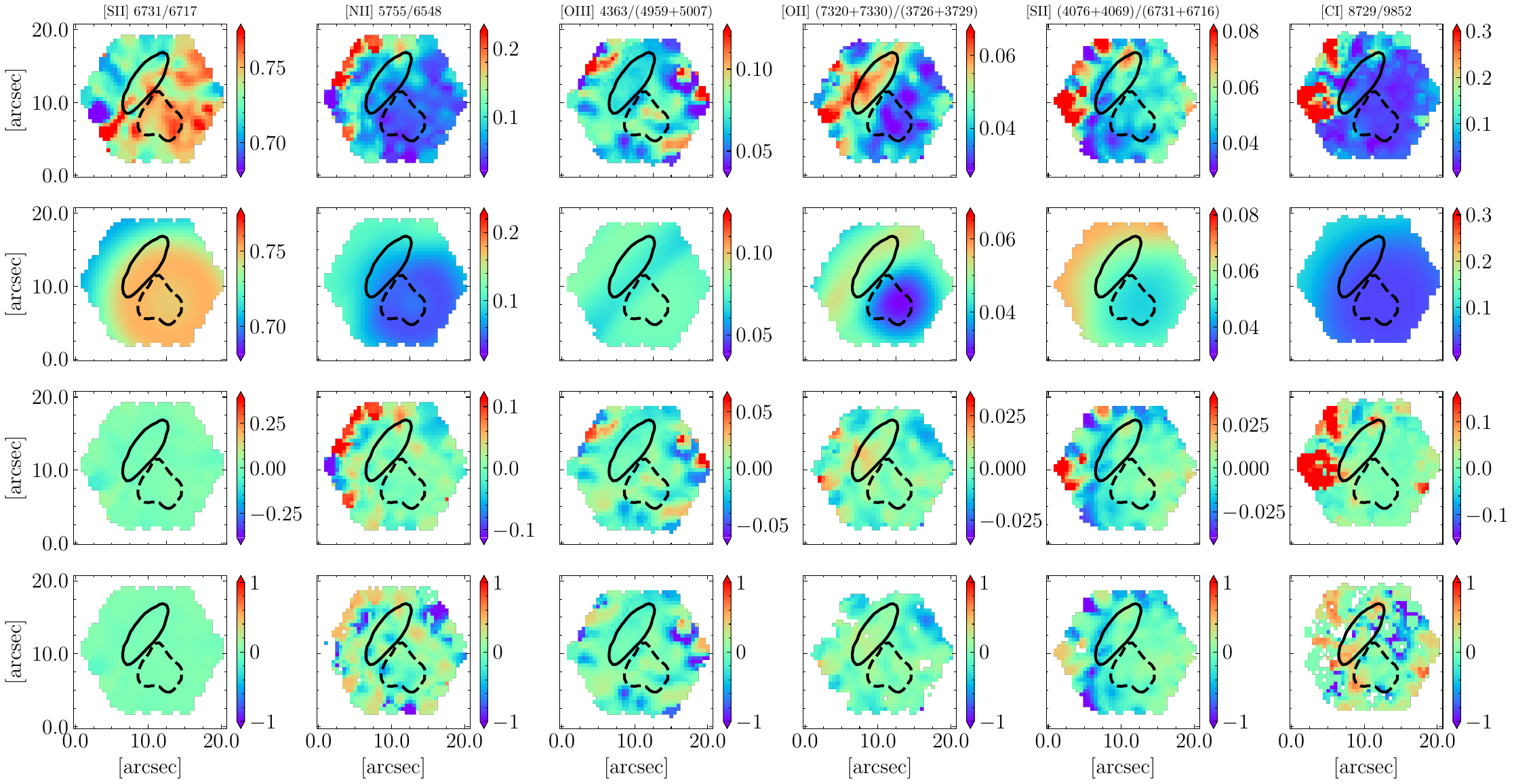}
	\caption{Shown in the top panels are the observed surface density maps 
		of six emission line ratios that are used to measure electronic density 
		and temperature maps in the previous figure (\autoref{fig:nT_map}). The predicted 
		maps of the same lines by Model 2 are shown in the panels in the 
		second row. The third and last rows show the absolute and relative 
		residual maps.}
	\label{fig:shock_model_lineratio}
\end{figure*}

From \autoref{fig:nT_map}, all the density maps present a prominent jump 
from the upper-left 
low density region ($\sim 1\ \mathrm{cm}^{-3}$) to the lower-right 
high density region ($\sim 100\ \mathrm{cm}^{-3}$), and the boundary 
of the two regions coincides with the location of the lane-like structure 
in the surface brightness maps of the [O{\sc iii}] $\lambda5007$-like group. 
This result qualitatively agrees with the predicted density profile in 
\autoref{fig:shock_model_g+s_ne_Te}.
Unlike $n_e$, the measurements of $T_e$ are different, depending on what temperature diagnostics are used. 
%The $T_e$ maps measured from [NII] 5755/6548 and [SII] (4076+4069)/(6731+6716) present no significant spatial structure, with a median temperature of $\log{(T_e/\mathrm{K})} \sim 4.0$. The $T_e$ map measured from [OII] (7320+7330)/(3726+3729) has a  slightly higher median temperature of $\log{(T_e/\mathrm{K})} \sim 4.2$, and a lane-like region of relatively high $T_e$ which coincides with the lane structure in the [OIII] brightness map. In contrast, [OIII] 4363/(4959+5007) results in much higher $T_e$, with a median of $\log{(T_e/\mathrm{K})} \sim 4.7$, and the $T_e$ map present a complex spatial distribution with several filamentary high-temperature structures. Finally, [CI] 8729/9852 gives rise to the lowest temperatures, $\log{(T_e/\mathrm{K})} \sim 3.9$, but the map shows a high-to-low $T_e$ jump from the upper-left to lower-right regions. 
In this case, only [O{\sc ii}] (7320+7330)/(3726+3729) captures the overall 
trend of the predicted temperature profile in \autoref{fig:shock_model_g+s_ne_Te}.
%In order to more clearly see the inconsistency between different diagnostic line ratios, in \autoref{fig:Te_ne_cross_plot} we show the constrained $T_e$-$n_e$ relations for two example spaxels: one located in the [OIII] lane and one from the H$\alpha$ clump. In both spaxels, it is clear that the $T_e$ constraints yielded by different line ratios are significantly inconsistent, spanning a range of two orders of magnitude at any given $n_e$. 
%\begin{figure}
%	\centering
%	\includegraphics[width=0.48\textwidth]{Te_ne_cross_plot.pdf}
%	\caption{The top-left panel is same as the top-right panel of Figure \ref{fig:large_scale_image}, but with the position of two example spaxels marked. The bottom-left panel is the constraints in electron temperature-density space given by different line ratios for spaxel (a). The line ratios corresponding to different colors are shown in the bottom-right panel. The solid line shows the constraints given by the value of line ratios, while the dash line with same color shows 1 $\sigma$ region caused by the uncertainty of line ratios. There is no blue and green dash line shown in bottom-left panel, because these the [NII] $\lambda$5755 and [CI] $\lambda$8729 are very weak in this spaxel, thus their 1 $\sigma$ regions cover the whole parameter space. The bottom-right panel is same as bottom-left panel but for spaxel (b).}
%	\label{fig:Te_ne_cross_plot}
%\end{figure}
However, we should note that it is not suitable to directly compare the profiles shown 
in \autoref{fig:shock_model_g+s_ne_Te} with the measurements in \autoref{fig:nT_map},
as the former are the intrinsic shock profiles without the Gaussian component
and observational effects (e.g. limited spatial resolution and projection effect) in the real data. 

In \autoref{fig:shock_model_lineratio} we compare the 
observed maps of the diagnosing line ratios (panels in the top row) with 
the predicted maps of the same line ratios from Model 2 in which we have included 
both the Gaussian component and the observational effects. Overall, our model 
can well reproduce the distribution of all the line ratios (the second row), although 
with subtle structures remaining in the maps of absolute residual (the third row) and 
relative residual (the bottom row). This result demonstrates that the discrepancies 
in $T_e$ as measured from different line ratios are caused by observational effects
as well as the contamination of non-shock components such as the Gaussian 
component in our case. It appears that different line ratios can be affected by 
these effects in different ways, thus leading to inconsistent measurements. 
We have examined the effect of the Gaussian component and the observational 
effects separately, and we find both effects to have contributed significantly
to the total discrepancies. Our analysis thus strongly warns that electronic density 
and temperature measurements based on observed emission line ratios can 
be seriously biased due to observational effects and non-shock components, 
and should be taken with caution.

\section{Discussion} \label{sec:discussion}

\subsection{The distance to CCN} \label{sec:distance}

If assumed to be a part of the Orion-Eridanus superbubble,
CCN should be at a distance ranging from $150\ \mathrm{pc}$ to 
$400\ \mathrm{pc}$, corresponding to the close and far ends of 
the superbubble
\citep{2016ApJ...827...42P, 2017ApJ...834..142K, 2017A&A...608A.148Z}.
Using the three-dimensional extinction map constructed by \citet{2022A&A...664A.174V}\footnote{https://explore-platform.eu/sda/g-tomo; v2: Resolution 10pc (sampling: 5$\mathrm{pc}$ - size: 3 x 3 x 0.8 $\mathrm{kpc}$)}, we find the differential extinction along the line of 
sight to CCN to present two peaks, located at $150\ \mathrm{pc}$ and 
$475\ \mathrm{pc}$, respectively. The two peaks coincide with the two 
edges of the Orion-Eridanus superbubble, and the CCN is likely to 
be associated with one of the peaks. In addition, we have attempted 
to constrain the distance to CCN with our data, by including the distance
as a free parameter in Model 2, $D_{\mathrm{CCN}}$. We obtain a 
best-fit distance of $D_{\mathrm{CCN}}=170\pm94\ \mathrm{pc}$.
This value is still consistent with the values 
from the first two considerations, given the large uncertainties. 
This result implies that the distance cannot be well constrained given 
the currently available data. Fortunately, we found that both the $B$ and $v_{\mathrm{sh}}$ have a weak correlation with the distance of CCN. On the other hand, although $n_{\mathrm{pre}}$ has a stronger correlation with $D_{\mathrm{CCN}}$, the marginalized uncertainty of $n_{\mathrm{pre}}$ only increases from $1.4\pm 0.2\ \mathrm{cm}^{-3}$ to $1.4\pm 0.3\ \mathrm{cm}^{-3}$.
The geometry parameters are indeed correlated 
with the distance, but our test shows that the data still prefer a small 
inclination angle, with models of $i < 30\ \mathrm{deg}$ showing 
a larger Bayesian evidence than those of $i \ge 30\ \mathrm{deg}$.

\subsection{The dust extinction in CCN} \label{sec:dust_extinction}

The dust extinction in CCN is also uncertain. An upper limit of the dust
extinction along the line of sight towards CCN can be estimated by 
integrating the differential extinction from \citet{2022A&A...664A.174V} 
over the distance range from 0 up to 475 $\mathrm{pc}$. This gives 
rise to $A_{5500 \text{\AA}}$ = 0.14 $\mathrm{mag}$, corresponding 
to a color excess of $\mathrm{E}(\mathrm{B}-\mathrm{V}) < 0.05$ mag
assuming the Milky Way extinction curve of \citet{CCM} with $\mathrm{R}_\mathrm{V} = 3.1$. This is smaller than the average value 
estimated from H$\alpha$ $\lambda$6563/H$\beta$ $\lambda$4861 in the MaNGA FoV,
$\mathrm{E}(\mathrm{B}-\mathrm{V})=0.12$ mag, using the 
electronic temperature for each spaxel as constrained by our 
best-fit model and assuming the Case B recombination coefficient.
We find that, however, different Balmer line ratios lead to inconsistent 
$\mathrm{E}(\mathrm{B}-\mathrm{V})$ values, implying that the 
extinction curve of \citet{CCM} may not be applicable for CCN,
or that the Case B assumption is not appropriate in shocked regions
(e.g. see Figure 4 of \citealt{Dopita2017_HHObject}). In fact, 
our best-fit shock model does predict higher H$\alpha$ $\lambda$6563/H$\beta$ $\lambda$4861 
ratios than that of Case B at fixed temperature. 
We have also attempted to include the dust extinction in our model. 
We find no significant improvement in goodness-of-fit, and the null hypothesis 
of ${E}(\mathrm{B}-\mathrm{V}) = 0$ can not be rejected at a
3$\sigma$ level. In summary, though without direct measurements 
of the dust content, constraints from different aspects consistently 
suggest that the dust extinction in CCN is rather limited. Our modelling 
presented in the previous section should have not been affected 
significantly by dust extinction.

\subsection{A single shock interacting with a cold gas cloud, or multiple shocks?}

The MaNGA data and our best-fit model suggest that the CCN region 
observed by MaNGA is ionized by a shock which moves across the 
MaNGA FoV from the lower-right corner to the upper-left corner. 
This direction is opposite to the shock direction indicated by the 
protruding double-arc structure of the whole CCN, which goes from 
left to right as suggested by Z97 and T07. A natural explanation on
this discrepancy is that CCN is a shocked cloud of cold gas 
(atomic gas, or molecular gas, or both). 
In this case, the overall curvature in the emission line surface brightness 
maps traces the shape of the gas cloud, 
but not the shock front. The protruding arcs to the right are actually the 
right edge of the cloud, which has collided head-on with the shock wave 
coming from the right. The shock has already swept into but not through
the cloud, and consequently the shock front is found on the left side of 
the arc and still within the cloud. In a recent theoretical study by 
\cite{2022MNRAS.513.3345K}, the simulation of a shocked H{\sc i} cloud 
produces a double-arc structure which is similar to CCN (see their Figure 4). 
Accordingly, one would expect to detect H{\sc i} on the left side of the 
double arc. We have examined the H{\sc i} data of the Galactic All Sky Survey 
\citep[GASS;][]{GASS1,GASS2}, finding no H{\sc i} enhancement 
at the position of CCN. Considering the low spatial resolution of GASS 
($14.4^{\prime}$), however, the H{\sc i} cloud may suffer from 
smearing effect and so the scenario of a shocked H{\sc i} cloud cannot 
be firmly ruled out with the current data. 
As abovementioned, we have also used the Purple Mountain Observatory 
(PMO) 13.7-m millimeter telescope to observe the $^{12}$CO(1-0) line in 
CCN (\autoref{sec:aux_data}). This observation provides an upper limit of  the H$_2$ surface 
density, which is $0.6\ M_\odot/\mathrm{pc}^2$ if we assume a conversion
factor of $\alpha_{\mathrm{CO}} = 4.35\ (M_\odot/\mathrm{pc}^2)/(\mathrm{K\ km/s})$ 
and a line width of $10\ \mathrm{km/s}$. This result largely rules out 
the scenario of a shocked molecular cloud, consistent with the 
simulation in \cite{2022MNRAS.513.3345K} where the CCN-like double 
arc can not be formed in a shocked H$_2$ cloud
(see their Figure 9).

%\subsection{Multiple shocks?} \label{sec:num_direc_shock}

The double-arc structure of the CCN could also be produced by multiple shocks,
rather than the interaction between a cold gas cloud and a single shock as 
discussed above. In fact, the idea of multiple shocks has its own 
supporting evidences, including significant differences in the line ratio 
distributions in/outside the ``clean region'' in \autoref{fig:clean_region_on_line_ratio},
residuals of the best-fit model in \autoref{fig:shock_model_g+s_residual}, 
the double-arc structure of the whole CCN, and the variation of shock velocity 
in different regions of CCN (e.g. $133\ \mathrm{km/s}$ in the MaNGA FoV studied 
in our work versus $40\ \mathrm{km/s}$ in the region studied by Z97). 
It's likely that there are two or even more shocks with different orientations in the 
CCN region. 

Both IFU observations covering the whole CCN and high-resolution 
H{\sc i} observations would be needed if one were to discriminate the scenario 
of gas cloud with a single shock and the picture of multiple shocks. 

\subsection{Source of the shock} \label{sec:num_direc_shock}

No matter which scenario is at work, as Z97 and T07 suggested, the source of 
the shock is most likely to be a supernova, which may be the one responsible for 
the Orion-Eridanus superbubble. In this case, however, the shock direction should 
go from left to right, thus in disfavor of the H{\sc i} cloud scenario. 
In addition, we failed to find any candidates of source supernova after having 
searched known supernovae and pulsars around CCN. 
Thus, the source of CCN is still an open question.

\subsection{The Gaussian component}\label{sec:discuss_gaussian}

\begin{figure}[t!]
	% \centering
	\includegraphics[width=0.45\textwidth]{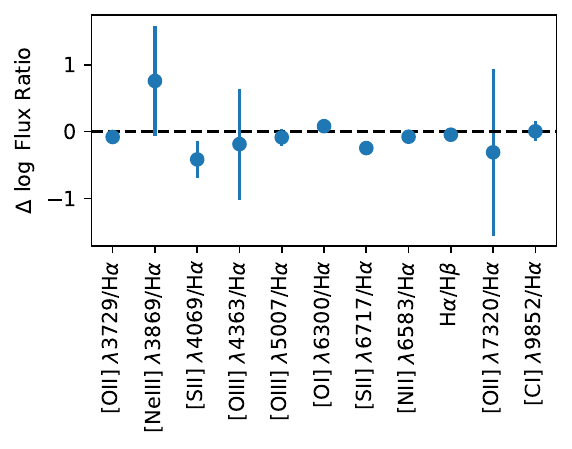}
	\caption{The residual of the best-fit integrated shock model obtained by comparing it with the line ratio of the Gaussian component. The error bar represents the 1 $\sigma$ uncertainty of the line ratio of the Gaussian component.}
	\label{fig:gaussian_shock_int}
\end{figure}

\begin{figure}[ht!]
	\centering
	\includegraphics[width=0.45\textwidth]{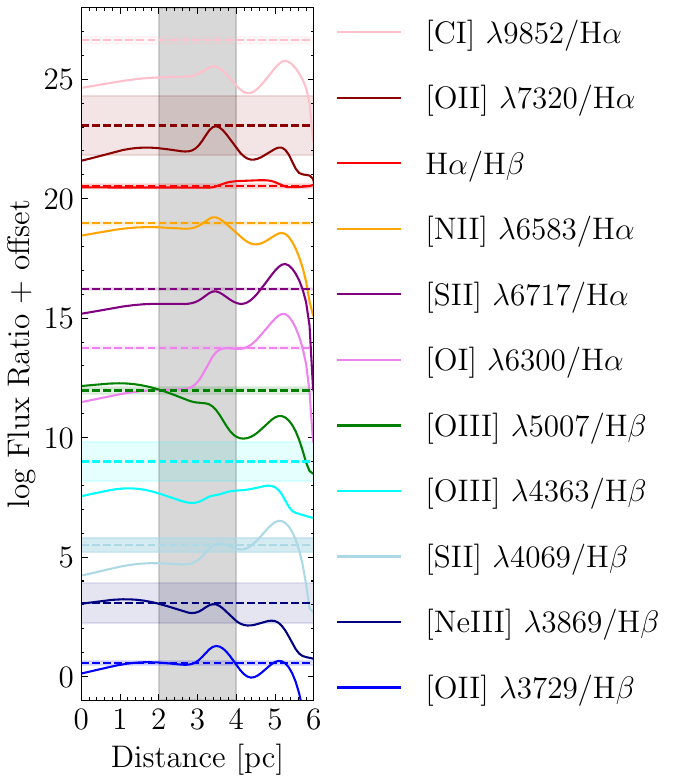}
	\caption{The line ratio profile of the precursor of Model 2 (solid line) is compared to the fitted line ratio of the Gaussian component (represented by the horizontal dashed line). The colorful shaded region represents the 1 $\sigma$ uncertainty of the fitted line ratio of the Gaussian component. Additionally, the grey shaded region indicates the coverage of the MaNGA FoV.}
	\label{fig:precursor_resolved}
\end{figure}

The physical origin of the Gaussian component is not immediately clear from our results. This component is possibly 
	caused by another resolved shock. We have attempted to fit the residual of the shock component in Model 2 with another 
	 shock moving in the same direction, 
	 i.e. from lower-right to upper-left in the MaNGA FoV.  We find that, however, some emission lines such as [S{\sc ii}] $\lambda$6717 and [N{\sc ii}] $\lambda$6583 cannot be fitted by this model. We have also tried with a shock direction from  upper-right to lower-left, which fails to fit the lines in Group II. Another possible case is that the second shock propagates along the line of sight. We thus use an integrated shock model (the default output of the {\tt MAPPINGS V}) to fit the line ratios in the Gaussian component. We find that a shock with $n_\mathrm{pre} = 0.1\ \mathrm{cm}^{-3}$, $v_\mathrm{sh} = 450\ \mathrm{km}/\mathrm{s}$, and $B = 0.1\ \mathrm{\mu G}$ can explain 
	 all the line ratios except [S{\sc ii}]$\lambda$6717/H$\alpha$ and 
	 [S{\sc ii}]$\lambda$4069/H$\alpha$, 
	 as can be seen from \autoref{fig:gaussian_shock_int}. 	Given the high velocity of this shock, we would expect the velocity field at the location 
	 of the Gaussian component to differ significantly from the other regions.
	 However, this is not observed in the MaNGA FoV (see
	 \autoref{fig:velocity_filed_good}). Therefore, another shock is unlikely to 
	 be the origin of the Gaussian component.

     Alternatively, the Gaussian component could be a clump of gas near the
     shocked region. In this scenario, the gas in this clump can be ionized by photons from shocked region. The most likely candidate that can be modeled 
     with {\tt MAPPINGS V} is the precursor of the shock in Model 2. 
     Although this gas clump may have a different geometry than the precursor, most of the line ratios of the Gaussian component agree with the 
     line ratio profiles of the shock precursor as predicted by Model 2 with 
     a precusor-to-shock front distance of 2-4 pc, as shown in 
     \autoref{fig:precursor_resolved}. Two of the line ratios, 
     [C{\sc i}] $\lambda$9852/H$\alpha$ $\lambda$6563 and [O{\sc iii}] 4363/H$\beta$ $\lambda$4861 show discrepancies between the 
     data and the model, which could be attributed to a variety of reasons,
     e.g. the homogenous gas assumption adopted in {\tt MAPPINGS V}, 
     differences in the geometry of this gas clump compared to the precursor component in {\tt MAPPINGS V}, differences in the metal abundance of the gas compared to our assumptions, or the contribution 
     of photons from the shocked region outside the MaNGA FoV. 
     However, investigation of these possibilities is beyond the scope of 
     the current paper. Considering that the majority of the line ratios 
     can be explained with no additional free parameters, we suggest that 
     the Gaussian component is very likely a clump of gas ionized by 
     photons from the shocked region, thus physically related to the 
     shocked region. Further studies in future are needed to pin down 
     this scenario.

\section{Summary} \label{sec:summary}

In this work, we have analyzed the integral field spectroscopy data for a small 
part of the Criss Cross Nebula (CCN), which was observed as a filler target 
by the MaNGA survey (\autoref{fig:large_scale_image}). 
At a distance of $\sim150-475$ pc, the spatial resolution 
of MaNGA ($\sim2.5^{\prime\prime}$) corresponds to a physical scale less than 
1000 AU. Within the MaNGA FoV, a hexagon covering 
$\sim20^{\prime\prime}\times20^{\prime\prime}$, we have obtained
the surface brightness map for 34 emission lines in the optical band, 
by fitting Gaussian profiles to the observed spectrum of each spaxel 
(\autoref{tab:lines} and \autoref{fig:lines_fitting_21_21}). 
We find the emission lines can be broadly divided into three groups
according to their surface brightness maps (\autoref{fig:surface_brightness_all}):
(1) the [O{\sc iii}] $\lambda$5007-like
group including seven high-ionization
lines and two [O{\sc ii}] auroral lines which uniformly present a remarkable 
lane of high brightness in the upper-left part of the MaNGA FoV, (2) the 
H$\alpha$ $\lambda$6563-like group including 23 low-ionization or recombination lines 
which mostly present a clump of high brightness in the lower-right part 
of the MaNGA FoV, and (3) the third group including only two lines, 
[O{\sc ii}] $\lambda$3726 and [O{\sc ii}] $\lambda$3729, which show clump-like 
structures at positions of both the [O{\sc iii}] $\lambda$5007 lane and the 
H$\alpha$ $\lambda$6563 clump. The velocity maps of different lines in a given group 
show high similarities, but vary from group to group (\autoref{fig:velocity_filed_good}).
The three categories of microstructures indicate both regularity and 
complexity in the CCN. 

We assume the lane structure in the [O{\sc iii}] $\lambda$5007-like group
is produced by a shock which is moving across the CCN from the 
lower-right to upper-left direction.  We use the measurements of 
emission line brightness and surface brightness ratios between different 
lines to constrain both physical parameters and geometry parameters 
of the shock model. We have examined two models: a simple model 
with a single shock plus a constant background (Model 1; \autoref{eqn:model_shock_profile}), 
and a more complex model with an additional Gaussian component (Model 2; \autoref{eqn:model_shock_profile2}). 
We show that the MaNGA data can be reasonably well-fitted with Model 2
(\autoref{fig:shock_model_1d_residual}; \autoref{fig:shock_model_g+s_residual}). 
The physical counterpart of the Gaussian component is likely another clump of gas ionized by photons from the shocked region (\autoref{fig:precursor_resolved}), though more data is needed to further confirm this scenario.
Our best-fit model parameters indicate that the shock has a pre-shock 
hydrogen density of $n_{\mathrm{pre}}=1.4\pm0.2$ cm$^{-3}$, a 
velocity of $133\pm5$ km/s, and a transverse magnetic field of 
 $B=1.6\pm0.3$ $\mu$G, with the shock direction orientated to the 
line of sight by an inclination angle of $19\pm14$ deg
(\autoref{tab:model_parameters}; \autoref{fig:geometry_model}). 

Finally, we have compared the electronic density and temperature profiles 
as predicted by Model 2 (\autoref{fig:shock_model_g+s_ne_Te})
with those calculated directly using observed emission line ratios as 
commonly-done in the literature (\autoref{fig:nT_map}). We find 
different line ratios to give rise to similar density maps but significantly
inconsistent temperature maps. The discrepancies can be attributed 
to observational effects caused by the limited spatial resolution 
of the data and the projection of the shock geometry, as well as 
the contamination of the additional Gaussian component. After 
including all these effects, our model can well reproduce all the 
emission line ratios that are relevant for calculating density and temperature
(\autoref{fig:shock_model_lineratio}). Thus, electronic densities 
and temperatures measured from observed line ratios should be taken 
with caution (see also \citealt{2022arXiv221014234C} for a similar 
caveat).

The CCN is an ideal lab for studying microstructures in shock regions 
and testing existing shock models. Considering its nearby distance, 
relatively high surface brightness and unique morphology, CCN will be 
a good template for studying similar nebulae in greater distance, 
e.g. those being/to be observed by the Local Volume Mapper of the SDSS V 
project \citep{SDSSV_LVM} and the AMAZE project 
\citep{AMAZE}. For the CCN itself, as mentioned,
future studies would benefit from high-resolution H{\sc i} and IFS 
observations, both covering the whole CCN. In fact, we have recently done
both observations, using MeerKAT and SITELLE on CFHT, and we expect 
to come back and present more scientific results soon.

\section*{Acknowledgments}
We thank the anonymous referee for his/her helpful comments.
We are grateful to Bringfried Stecklum for providing the [S{\sc ii}] and H$\alpha$
narrow band images used in T07, Christophe Morisset and Alexandre Alarie for 
help on calculating spatial resolved shock profiles with the 
{\tt MAPPINGS V} code, and the staff at Delingha Station of the PMO for 
help with observations and data reduction. 
This work is supported by the National Key R\&D Program of China
(grant NO. 2022YFA1602902), and the National Science
Foundation of China (grant Nos. 11821303, 11733002, 11973030,
11673015, 11733004, 11761131004, 11761141012).
RY acknowledges support by the Hong Kong Global STEM Scholar Scheme, by the Hong Kong Jockey Club through the JC STEM Lab of Astronomical Instrumentation program, and by a grant from the Research Grant Council of the Hong Kong (Project No. 14302522). CC is supported by the National Natural Science Foundation of China, No. 11933003, 12173045. This work is sponsored (in part) by the Chinese Academy of Sciences (CAS), through a grant to the CAS South America Center for Astronomy (CASSACA). We acknowledge the science research grants from the China Manned Space Project with NO. CMS-CSST-2021-A05.

Funding for the Sloan Digital Sky Survey IV has been provided by the Alfred P. Sloan Foundation, the U.S. Department of Energy Office of Science, and the Participating Institutions. 
SDSS-IV acknowledges support and resources from the Center for High Performance Computing at the University of Utah. The SDSS website is www.sdss.org.

SDSS-IV is managed by the Astrophysical Research Consortium for the Participating Institutions of the SDSS Collaboration including the Brazilian Participation Group, the Carnegie Institution for Science, Carnegie Mellon University, Center for Astrophysics | Harvard \& Smithsonian, the Chilean Participation Group, the French Participation Group, Instituto de Astrof\'isica de Canarias, The Johns Hopkins University, Kavli Institute for the Physics and Mathematics of the Universe (IPMU) / University of Tokyo, the Korean Participation Group, Lawrence Berkeley National Laboratory, Leibniz Institut f\"ur Astrophysik 
Potsdam (AIP), Max-Planck-Institut f\"ur Astronomie (MPIA Heidelberg), Max-Planck-Institut f\"ur 
Astrophysik (MPA Garching), Max-Planck-Institut f\"ur Extraterrestrische Physik (MPE), National Astronomical Observatories of China, New Mexico State University, New York University, University of Notre Dame, Observat\'ario Nacional / MCTI, The Ohio State University, Pennsylvania State 
University, Shanghai Astronomical Observatory, United Kingdom Participation Group, 
Universidad Nacional Aut\'onoma de M\'exico, University of Arizona, University of Colorado Boulder, University of Oxford, University of Portsmouth, University of Utah, University of Virginia, University of Washington, University of Wisconsin, Vanderbilt University, and Yale University.

This research has made use of the Millimeter Wave Radio Astronomy Database.
This research has used data, tools or materials developed as part of the EXPLORE project that has received funding from the European Union's Horizon 2020 research and innovation programme under grant agreement No 101004214. 

The authors acknowledge the Tsinghua Astrophysics High-Performance Computing platform at Tsinghua University for providing computational and data storage resources that have contributed to the research results reported within this paper.


\vspace{5mm}
\facilities{MaNGA, GASS, Purple Mountain Observatory (PMO) 13.7-m millimeter telescope, 2 m telescope in Thüringer Landessternwarte Tautenburg (Germany), China Near Earth Object Survey Telescope}


\software{Python3\citep{python3}, NumPy\citep{numpy}, Matplotlib\citep{matplotlib}, Joblib\citep{joblib}, Astropy\citep{astropy:2013, astropy:2018, astropy:2022}, SciPy\citep{scipy}, statsmodels\citep{statsmodels}, pandas\citep{pandas_1, pandas_2}, tqdm\citep{tqdm}, LMFIT\citep{lmfit}, PyNeb\citep{pyneb}, MAPPINGS V\citep{MAPPINGSV}}



% \appendix


\bibliography{sample631}{}
\bibliographystyle{aasjournal}



\end{document}